\documentclass[11pt,twoside]{article}

\usepackage{indentfirst}
\usepackage{amssymb}
\usepackage[utf8]{inputenc}
\usepackage{ae}
\usepackage{amsthm,amsfonts,latexsym,fancyhdr,amsmath,tabularx}
\usepackage{graphicx}
\usepackage{subfigure, multicol}
\usepackage{epsfig}
\usepackage{verbatim}
\usepackage{pgf,tikz,pgfplots}
\usetikzlibrary{arrows, shapes, trees}
\usetikzlibrary{calc}
\pgfplotsset{compat=1.14}
\usepackage{lineno,hyperref}
\usepackage{expdlist}

\newenvironment{singlespace}{ \small
\normalsize}{ \small
\normalsize}

\newtheorem{theorem}{Theorem}[section]

\newtheorem{conjectura}{Conjecture}[section]

\newtheorem{definition}{Definition}[section]
\newtheorem{example}{Example}[section]

\newtheorem{proposition}{Proposition}[section]
\newtheorem{remark}{Remark}[section]

\begin{document}

\pagestyle{fancy}

\begin{singlespace}

\title{Construction of Hyperbolic Signal Sets from the Uniformization of Hyperelliptic Curves}
\author{Érika Patricia Dantas de Oliveira Guazzi\thanks{The author is with the Departamento de Matemática, 	Universidade Tecnológica do Paraná, Campo Mourão, PR, Brazil e-mail: erikapatricia@utfpr.edu.br}, ~ and ~ Reginaldo Palazzo Jr.\thanks{The author is with the Departamento de Comunicações, FEEC-UNICAMP. This work has been supported by FAPESP, CNPq, and CAPES, Brazil. email:palazzo@dt.fee.unicamp.br}}
\date{}

\maketitle	

\begin{abstract}
In this paper, we present a new approach to the problem of designing hyperbolic signal sets matched to groups by use of Whittaker's proposal in the uniformization of hyperelliptic curves via Fuchsian differential equations (FDEs). This systematic process consists of the steps: 1) Obtaining the genus, $g$, by embedding a discrete memoryless channel (DMC) on a Riemann surface; 2) Select a set of symmetric points in the Poincaré disk to establish the hyperelliptic curve; 3) The Fuchsian group uniformizing region comes by the use of the FDE; 4) Quotients of the FDE linearly independent solutions, give rise to the generators of the associated Fuchsian group. Equivalently, this implies the determination of the decision region (Voronoi region) of a digital signal. Hence, the following results are achieved: 1) from the solutions of the FDE, the Fuchsian group generators are established. Since the vertices of the fundamental polygon are at the boundary of $\mathbb{D}^2$, its area (largest possible) implies the least symbol error probability as a performance measure of a digital communication system; 2) a relation between the parameters of the tessellation $\{p,q\}$ and the degree of the hyperelliptic curve is established. Knowing $g$, related to the hyperelliptic curve degree, and $p$, number of sides of the fundamental polygon derived from Whittaker's uniformizing procedure, the value of $q$ is obtained from the Euler characteristic leading to one of the $\{4g,4g\}$ or $\{4g+2, 2g+1\}$ or $\{12g-6,3\}$ tessellation. These tessellations are important due to their rich geometric and algebraic structures, both required in classical and quantum coding theory applications.
\end{abstract}

\end{singlespace}

\textbf{keywords}: Fuchsian groups, Uniformization, Fuchsian differential equation, Hyperbolic plane

\newpage

\section{Introduction}

The proposal to use topological spaces as a first step towards the design of general systems (pattern recognition, and biological, just to name a few)  is getting the attention lately, by showing its efficiency in the construction of new signal constellations, \cite{chacon2, chacon1, vandenberg2, edsonb, JAA}, in the construction of new topological quantum codes, \cite{kitaev, dennis, bombin, preskill, delgado}, in the construction of holographic codes, \cite{pasta, latorre, brill}, and several other problems as relevant as the ones just mentioned. For instance, regarding the design of reliable, efficient, and less computationally complex digital communication systems, this change of paradigm makes it possible to include new mathematical concepts, like the ones shown in this paper, not previously considered in the aforementioned systems.

It is by the attributed relevance to the communication channel that it will be possible to introduce new strategies to overcome interferences, noise, and other forms of impairments. In this direction, consider a discrete memoryless channel (DMC). The identification of the proper topology follows from the embedding of this channel on compact surfaces, minimal surfaces, \cite{joaodeus2002}, or composition of both as is the case of holographic codes, \cite{pasta}.

Among the topological invariants, the \textbf{genus}, $g$, of a surface is the relevant invariant in this new proposal, for it provides the possibility to analyze the surface's geometric and algebraic properties where the DMCs will be embedded, \cite{Rodrigo, Stillwell2012, Firby}.

Since our interest lies on the cases where $g\ge 2$, it follows that subgroups and quotient of groups associated to the discrete groups of isometries (Fuchsian groups) of a regular tessellation in the hyperbolic plane of the type $\{p,q\}$ establishes the signal set that will be constructed, \cite{Rodrigo, lazari}, where $p$ denotes the number of sides of the regular polygon and $q$ denotes the number of such polygons meeting at each vertex. The strategy to systematize the procedure is to interpret the fundamental region of the tessellation $\{p,q\}$ as the uniformizing region of an algebraic curve derived from the Fuchsian differential equation corresponding to the compact surface, which embeds the DMC channel.

In \cite{Whittaker1898, Whittaker1929, Whittaker1930, Mursi1930, burnside1893, Girondo2003, Brezhnev2008, Brezhnev2009} the relationship between the Fuchsian groups and the Fuchsian differential equations is considered. In \cite{Anderson2017}, it was realized a detailed study of these FDEs (where the hypergeometric differential equations stand out as the most analyzed ones) and the relationship of these equations with the elements of the hyperbolic geometry. In \cite{Whittaker1929}, the objective was to establish the necessary and sufficient conditions such that the quotient of two FDE solutions should be Fuchsian.

The objective of this paper is to propose a systematic approach for the construction of hyperbolic signal sets from the uniformization of hyperelliptic curves by the use of the Fuchsian differential equation. To the best of our knowledge, the procedure leading to the determination of the uniformizing region of a hyperelliptic curve has not been employed previously in problems related to communication theory. 

To achieve this goal, the steps to be followed in this systematic approach are described next:

\begin{itemize}
	\item[1) ] Let $C$ be a DMC and $K_{m,n}$ the corresponding complete bipartite graph;
	
	\item[2) ] Determine the minimum, $g_{m}$, and maximum, $g_{M}$, values taken by $g$ ($g_{m} \le g \le g_{M}$) of the compact surface where $K_{m,n}$ may be embedded as a two-cell embedding, \cite{Ringeisen1972}, that is,
	\[g_m(K_{m,n})=\left\{\frac{(m-2)(n-2)}{4}\right\}\]
	for $m, n \geq 2$, and $\{a\}$ denotes the least integer greater than or equal to the real number $a$, and
	\[g_M(K_{m,n})=\left[\frac{(m-1)(n-1)}{2}\right]\]
	for $m, n \geq 1$, and $[a]$ denotes the greatest integer less than or equal to the real number $a$.
	For more detailed information, we refer the reader to \cite{Rodrigo}, \cite{Ringeisen1972} and \cite{Ringel1965}.
	
	\item[3) ] Knowing $g_m$ and $g_M$, choose a set of symmetric points $S=\{s_1, s_2, \ldots, s_{(2g+1)}\}$ in the Poincaré disk and for each value of $g$ such that $g_m\leq g \leq g_M$, establish the hyperelliptic curve $y^2=f(z)=(z-s_1)(z-s_2)\ldots(z-s_{(2g+1)})$;
	
	\item[4) ] Given the second order FDE \[\frac{d^2y}{dz^2}+\frac{3}{16}\left[\left(\frac{f'(z)}{f(z)}\right)^2-\frac{2g+2}{2g+1} \frac{f''(z)}{f(z)}\right]y=0\] determine the fundamental region which will uniformize the hyperelliptic curve, or equivalently, determine the associated Fuchsian group, where the generators of the group are quotients of the linearly independent solutions of this FDE.
\end{itemize}

The Whittaker proposal can not be generalized to any algebraic curve. To the best of our knowledge, Rankin \cite{Rankin1958} establishes a set with the largest number of algebraic curves for which the Whittaker proposal holds, that is, for the curves of the form $y^m=z^p(z^n-1)^r$, where $m$, $p$, $n$ and $r$ are positive integers. It has been shown in \cite{Girondo2003} that although the established Rankin algebraic curves have many automorphisms, however there are algebraic curves with less automorphisms satisfying Whittaker proposal. We call attention to the fact that our interest is in hyperelliptic curves where $m=2$, $p=0$, $n=2g+1$ or $n=2g+2$, and $r=1$, implying that $y^2 = z^{2g+1} \pm 1$ or $y^2 = z^{2g+2} \pm 1$. The relevance of these two curves lies on the fact that they lead to the $\{4g,4g\}$ and $\{4g+2,2g+1\}$ tessellations, respectively, for which there is an algorithm to find arithmetic Fuchsian groups, \cite{cintya}, leading to the construction of hyperbolic lattice codes. We consider in this paper only the case $n=2g+1$ since the case $n=2g+2$ has the same number of branch points.

Hence, given a hyperelliptic curve satisfying Whittaker's proposal with all of its roots at the boundary of the Poincaré disk, the following results are expected to be achieved.

\begin{itemize}
	\item To identify the generators of the Fuchsian subgroup associated with the fundamental region which uniformizes the given hyperelliptic curve;
	
	\item To establish a relation between the degree of the hyperelliptic curve and the parameters $p$ and $q$ of the tessellation $\{p,q\}$, where $p$ is derived from the uniformizing region of the hyperelliptic curve.
\end{itemize}

This paper is organized as follows. In Section \ref{cap_revisao}, the necessary concepts for grasping the content of the paper is presented. In Section \ref{Desenvolvimento_Whittaker}, Whittaker's proposal, \cite{Whittaker1929}, is briefly reviewed and the curve $y^2 = z^5 +1$ is taken into consideration. In Section \ref{cap_Primeiros_Resultados}, we show how to obtain the Fuchsian subgroup associated with a given hyperelliptic curve whose fundamental region is the limiting case, that is, all of its roots (vertices) are at the boundary of $\mathbb{D}^2$. A relation between the hyperelliptic curve degree and the parameters of the tessellation is also derived. Finally, in Section \ref{conclusoes}, the conclusions are drawn. In the Appendix, it is considered a detailed description of Whittaker's proposal in order to fill in some essential steps left out of in obtaining the generators of the Fuchsian group associated with the uniformizing region of the given hyperelliptic curve.

\section{Preliminaries} \label{cap_revisao}

In this section, we present some essential concepts needed in this paper.

\subsection{Elements of hyperbolic geometry}

When considering surfaces with genus $g\geq 2$, the corresponding geometry is the \textbf{hyperbolic geometry}. There are four models in this geometry. However, we consider only the model of the Poincaré disk $\mathbb{D}^2=\{z \in \mathbb{C}||z|<1\}$, we refer the reader to \cite{Katok1992} for more detailed information.

\begin{definition} \cite{Katok1992}
	The transformations identified in $PSL(2,\mathbb{R})=\linebreak SL(2,\mathbb{R})/\{\pm I\}$ are classified in three types according to the value taken by the absolute value of the matrix trace. Let $T(z)=\frac{az+b}{cz+d}$, $a,b,c,d \in \mathbb{Z}$, with $ad-bc=1$. In this way, $T$ is an elliptic transformation, if $|Tr(T)|=|a+d|<2$, a parabolic transformation, if $|Tr(T)|=|a+d|=2$, and a hyperbolic transformation, if $|Tr(T)|=|a+d|>2$.
\end{definition}

Given a set $A\subset \mathbb{D}^2$, its hyperbolic area $\mu(A)$, is defined as
\[
\mu(A)=\int _A \frac{4dxdy}{(1-(x^2 + y^2))^2},
\]
if the integral exists and is finite.

\begin{theorem}\cite{Walkden}(Gauss-Bonnet)
	Let $A$ be a hyperbolic triangle with angles $\alpha$, $\beta$, and $\gamma$. Then, the hyperbolic area of $A$ is given by $\mu(A)=\pi -(\alpha + \beta + \gamma)$.
\end{theorem}

\begin{definition} \cite{Katok1992}
	The Mobius transformations are isometries, that form a subgroup of the isometries of the Poincaré disk, $Isom(\mathbb{D}^2)$.
\end{definition}

\begin{definition} \cite{Katok1992}
	A regular tessellation of the hyperbolic plane is a partition consisting of polygons, all congruent, subject to the constraint of intercepting only at sides and vertices, and also having the same number of polygons meeting at each vertex, independent of the vertex. Therefore, there are infinite regular tessellations in $\mathbb{D}^2$.
\end{definition}

\begin{definition} \cite{Katok1992}
	The side-pairing of a regular hyperbolic polygon $P_p$ is a set of isometries $\phi=\{T_{\tau}|\tau \in \mathcal{A}\}$ such that, for each side $\tau \in \mathcal{A}: $ there exists a side $\tau^{'}=\tau $ and the isometries $T_{\tau}$ and $T_{\tau^{'}}$, satisfying $T_{\tau^{'}}=T_{\tau}^{-1}$ and if $\tau$ is a side of $P_p$ then $\tau^{'}=P\cap T_{\tau}^{-1}(P)$.
\end{definition}

\begin{theorem}\cite{Stillwell2012}\label{theorem11}
	Let $P_p$ be a Dirichlet region associated with the Fuchsian group $\Gamma _{p}$ and $v_1, \cdots , v_p$ the $p$ vertices of $P_p$. Let $v_1, \cdots , v_t$, where $t\le p$, be the vertices of a cycle and $\delta_1, \cdots , \delta_t$ the internal angles in the corresponding vertices. If $m$ denotes the order of the stabilizer in $\Gamma _p$ of one of the vertices of the cycle, then
	\begin{equation}\label{eqaa}
	\delta_1 + \delta _2 + \cdots + \delta _t = \frac{2\pi}{m}.
	\end{equation}
\end{theorem}

As a consequence of Theorem \ref{theorem11}, it follows that if one of the vertices $v_1, \cdots , v_t$, $t\le p$, of a cycle of vertices is not a fixed point, then $m=1$. Since the Fuchsian groups $\Gamma _p$ coming from the uniformization process for the hyperelliptic curve does not have elliptic elements, it follows that the condition in (\ref{eqaa}) is given by,
\[
\delta _1 + \delta _2 + \cdots + \delta _t = 2\pi,
\]
where $t\le p$. Furthermore, given a regular tessellation $\{p,q\}$, since there are $q$ of these polygons meeting at each vertex of $P_p$, it follows that the internal angles in the corresponding vertices are $2\pi/q$. Hence, by Theorem \ref{theorem11}, it follows that each cycle must have exactly $q$ vertices and the amount of cycles in a tessellation $\{p,q\}$ is $\frac{p}{q}$. Thus, $q$ must divide $p$.

\subsection{Cross-ratio}\label{def_razao_cruzada}

\begin{definition}\cite{Katok1992}
	Let $z_1, z_2, z_3, z_4 $ be distinct points in the extended complex plane. Thus, the cross-ratio $[z_1, z_2, z_3, z_4]$ is defined by
	\[
	[z_1, z_2, z_3, z_4]= \frac{(z_1 - z_2)(z_3 - z_4)}{(z_2 - z_3)(z_4 - z_1)}.
	\]
\end{definition}

\subsection{Differential equations in the complex plane}\label{sec_ed_complexa}

In this subsection, we present the concepts and results related to second-order Fuchsian differential equations (FDEs).

\subsubsection{Fuchsian differential equations}

Differential equations where all the singular points are regular singular points are called \textbf{Fuchsian differential equations} or, simply, \textbf{Fuchsian equations}.

\begin{definition}\cite{Sotomayor1979}
	The second order differential equation
	\begin{equation}\label{eq_rev_edf}
	y''(z)+ p(z)y'(z)+ q(z)y(z ) = 0,
	\end{equation}
	is called second order Fuchsian differential equation if every singular point in the extended complex plane is regular, that is, if the singularity in $p(z)$ is a simple pole and in $q(z)$ is a pole of at most order 2.
\end{definition}

\begin{example}
	The hypergeometric differential equation (HDE) is given by
	\[ z\left( z - 1\right) y''(z) + \left[ \left( \alpha + \beta + 1 \right) z - \gamma \right] y'(z) + \alpha \beta y(z) = 0, \]
	and has three regular singular points at $z = 0$, $z = 1$ and $z = \infty$.
\end{example}

The solution of a second-order FDE with at least three regular singular points may be reduced to a solution of an HDE. The solutions of the FDE are established using the Frobenius method, \cite{Sotomayor1979}. This method makes it possible to calculate the solutions of an $n$ order FDE in a regular singular point.

\subsection{Solutions to the Hypergeometric differential equation} \label{solucaohipergeometrica}

In this section we present the solutions of the hypergeometric differential equation related to each one of the three singularities at $z=0$, $z=1$ and $z=\infty$, for more detailed information we refer the reader to \cite{Sotomayor1979}, \cite{kristensson2010}, \cite{Bateman1953a}. The reason for presenting the HDE solutions is that they are part of the final steps toward the solution of the FDE.

Consider the hyperelliptic differential equation
\begin{equation}\label{eqhiper}
z(1-z)y''(z)+[\gamma-(1+\alpha+\beta)z]y'(z)-\alpha \beta y(z)=0.
\end{equation}
From \cite{Sotomayor1979}, the solutions of the equation (\ref{eqhiper}) at the singularities are given by:
\begin{enumerate}
	\item Solutions at $z=0$: $\gamma \neq 0$ and $\gamma \not\in \mathbb{Z}$ negative.\\
	$w_{10}(z)=F(\alpha, \beta; \gamma; z)$\\
	$w_{20}(z)=z^{1-\gamma} F(\alpha +1 -\gamma, \beta+1 -\gamma; 2-\gamma; z)$\\
	\item Solutions at $z=1$: $\gamma -(\alpha+\beta) \not \in \mathbb{Z}$ positive.\\
	$w_{11}(z)=F(\alpha, \beta; 1-\gamma+\alpha + \beta; 1-z)$\\
	$w_{21}(z)=(1-z)^{\gamma-\alpha - \beta} F(\gamma-\beta, \gamma-\alpha; 1 -\gamma-\alpha- \beta; 1-z)$\\
	\item Solutions at $z= \infty$: $\alpha-\beta \not \in \mathbb{Z}$ negative.\\
	$w_{1\infty}(z)= z^{-\alpha} F(\alpha, 1-\gamma+\alpha; 1- \beta+\alpha; z^{-1})$\\
	$w_{2\infty}(z)=z^{- \beta} F(\beta, 1-\gamma-\beta; 1 +\beta-\alpha; z^{-1})$\\
\end{enumerate}

\begin{definition}\cite{Bateman1953a}
	The hypergeometric function is defined as
	$$F(\alpha, \beta; \gamma; z)=\frac{\Gamma(\gamma)}{\Gamma(\alpha) \Gamma(\beta)}\sum_{n=0}^{\infty} \frac{\Gamma(\alpha +n)\Gamma(\beta +n)}{\Gamma(\gamma +n)} \frac{z^n}{n!}.$$
\end{definition}

Since the solutions are given by expressions that involve the hypergeometric function, we present some of its properties.
\begin{enumerate}
	\item $F(\alpha, \beta; \gamma; z)=F(\beta, \alpha; \gamma; z)$
	\item $F(\alpha, \beta; \gamma; 1)=\frac{\Gamma(\gamma)\Gamma(\gamma-\alpha-\beta)}{\Gamma(\gamma-\alpha) \Gamma(\gamma-\beta)}$, $\gamma \neq 0,-1,-2, \ldots$ e $ Re(\gamma)>Re(\alpha+\beta)$
	\item $F(\alpha, \beta; \gamma; 0)=1$
	\item $F(\alpha, \beta; \beta; z)=(1-z)^{-\alpha}$
	\item $F(\alpha, \beta; \gamma; z)=(1-z)^{\gamma-\alpha-\beta} F(\gamma-\alpha, \gamma-\beta; \gamma; z)$
	\item $(1-z) F(\alpha, \beta; \gamma -1; z)=1+z\left(\frac{\alpha+\beta-2\gamma+1}{\gamma-1}\right) F(\alpha, \beta; \gamma; z)+ \frac{(\alpha-\gamma)(\beta-\gamma)}{\gamma(\gamma-1)} F(\alpha, \beta; \gamma +1; z)$.
\end{enumerate}

The analytical continuation, \cite{Bateman1953a}, relates the solutions of the HDE. Thus, the solutions at $z=0$ may be rewritten as the linear combination of the solutions at $z=1$, namely,

\begin{eqnarray*}
	F(\alpha, \beta; \gamma; z) & = & \frac{\Gamma(\gamma)\Gamma(\gamma-\alpha-\beta)}{\Gamma(\gamma-\alpha) \Gamma(\gamma-\beta)} F(\alpha, \beta; \alpha+\beta-\gamma +1; 1-z) \\
	& + & \frac{\Gamma(\gamma)\Gamma(\alpha+\beta-\gamma)}{\Gamma(\alpha) \Gamma(\beta)} (1-z)^{\gamma-\alpha-\beta} F(\gamma-\alpha, \gamma-\beta; \gamma-\alpha-\beta +1; 1-z)
\end{eqnarray*}
with $|arg(1-z)|<\pi$.

Besides, the solutions at $z=0$ may also be rewritten as the linear combination of the solutions at $z=\infty$, namely,
\begin{eqnarray*}
	F(\alpha, \beta; \gamma; z) & = & \frac{\Gamma(\gamma)\Gamma(\beta-\alpha)}{\Gamma(\gamma-\alpha) \Gamma(\beta)} (-z)^{-\alpha} F(\alpha, 1-\gamma+\alpha; 1-\beta+\alpha; z^{-1}) \\
	& + & \frac{\Gamma(\gamma)\Gamma(\alpha-\beta)}{\Gamma(\alpha) \Gamma(\gamma-\beta)} (-z)^{-\beta} F(\beta, 1-\gamma+\beta; 1-\alpha+\beta; z^{-1})
\end{eqnarray*}
with $|arg(-z)|<\pi$.

\begin{remark}
	The development of $(-x)^{-k}$ makes use of the fact that $\ln(-1)=\pm \pi i$:
	\begin{eqnarray*}
		(-x)^{-k} & = & e^{\ln(-x)^{-k}}=e^{-k\ln(-x)}=e^{-k\ln((-1)x)}=e^{-k[\ln(-1)+\ln(x)]} \\
		& = & e^{-k\ln(-1)} e^{-k\ln(x)}=e^{\pm k\pi i} e^{\ln(x)^{-k}}\\
		& = & x^{-k} e^{\pm k \pi i}.
	\end{eqnarray*}
\end{remark}

\subsection{Uniformization}

This subsection follows a similar concept in \cite{Ford} and \cite{rocha} closely, due to the didactical and clear form of exposition. Consider the circle given by $X^2+ Y^2=1$. The two parametric forms associated with this equation are given by $Y=\sin z$, $X=\cos z$, and by $Y=\frac{2z}{1+z^2}$ and $X=\frac{1-z^2}{1+z^2}$. Note that $X$ is a two-valued function of $Y$, whereas, in the two parametric forms, both $Y$ and $X$ are one-valued functions. Therefore, the parametric solutions uniformize the circle. These functions are called \textit{automorphic functions}.

From the Riemann surface theory it is known that a compact Riemann surface can be viewed either as an algebraic curve $G(X,Y) = 0$, or as the quotient space of the unit disc $\mathbb{D}^2$ by the action of a Fuchsian group. Hence, let $f_1(z)$, $f_2(z)$ be two automorphic functions belonging to the same group and having the same definition domain. Thus, there exists an algebraic relation $G(f_1,f_2)=0$. If $X$ is an algebraic function of $Y$, defined by the relation $G(X,Y)=0$, then the functions $X=f_1(z)$ and $Y=f_2(z)$ uniformize the algebraic function. Note that given $G(X,Y)=p_0(Y)X^m + p_1(Y)X^{m-1} + \cdots + p_m(Y)=0$, where $p_i(Y)$ are polynomials, and $G(X,Y)$ is irreducible, then for each value of $Y$ there are $m$ distinct values of $X$. Then, the Riemann surface of $X$ as a function of $Y$ is a closed surface of two sides with $m$-sheets with a finite number of ramification points.

\subsection{Hyperelliptic functions}\label{hc}

We refer the reader to \cite{Ford}, \cite{Whittaker1898}, and \cite{shaska2019}, and references within, for more detailed information.

\begin{definition}
Let $K$ be a field and $\overline{K}$ its algebraic closure. A hyperelliptic curve $C$ of genus $g$, $(g\geq 1)$, over $K$ is an equation of the form:
\begin{equation}
C: y^2+h(z)y=f(z) \quad \mbox{in} \quad K[z,y],
\end{equation}
where $h(z) \in K[z]$ is a polynomial of degree at most $g$, $f(z) \in K[z]$ is a monic polynomial of degree $2g+1$, and do not exist solutions $(z,y) \in \overline{K} \times \overline{K}$ simultaneously satisfying the equation $y^2+h(z)y=f(z)$ and the partial derivatives equations $2y+h(z)=0$ and $h^{\prime}(z)y-f^{\prime}(z)=0$. A singular point in $C$ is a solution $(z,y) \in \overline{K} \times \overline{K}$ simultaneously satisfying the equation $y^2+h(z)y=f(z)$ and the partial derivatives equations $2y+h(z)=0$ and $h^{\prime}(z)y-f^{\prime}(z)=0$.
\end{definition}

Hyperelliptic curves are a special class of algebraic curves and may be seen as a generalization of elliptic curves. There are hyperelliptic curves for every genus $g\geq 1$. The degree of the polynomial determines the genus of the curve: a polynomial of degree $2g+1$ or $2g+2$ implies that the genus of the curve is $g$. 

\section{Whittaker's Proposal for Determining the Uniformization Region of a Hyperelliptic Curve}\label{Desenvolvimento_Whittaker}

The composition of a signal set in the modulator and a set of decision rules in the demodulator establishes a discrete memoryless channel. Hence, this channel may be represented by a graph. It is by the embedding of this graph on surfaces that the genus, $g$, is obtained, \cite{joaodeus2002} and \cite{Ringeisen1972}. Knowing $g$, the degree of the hyperelliptic curve is either $2g+1$ or $2g+2$, for $g\ge 1$. Knowing the linearly independent solutions to the Fuchsian differential equation (determination of the generators of the Fuchsian group), the uniformization region of the hyperelliptic curve is known. Therefore, the number of sides of the fundamental polygon is known. From Theorem \ref{theorem11}, it follows that each cycle must contain exactly $q$ vertices and the number of cycles in a tessellation $\{p,q\}$ is given by $p/q$. Consequently, this ratio provides the number of vertices $V$ of the fundamental polygon to be used in the Euler characteristic of the surface, that is, $\chi (\Gamma_p)=V-E+F=p/q \,- \, p/2 \, +\, 1=2-2g$, where $E$ denotes the number of sides, and $F$ the number of faces. However, knowing the genus, $g$, and the number of sides, $p$, from the Euler characteristic, the value of $q$ is found. Therefore, the tessellation $\{p,q\}$, is specified. Note that these values have to satisfy the inequality $(p-2)(q-2)> 4$, \cite{Firby}. The previous arguments reveal the importance of the uniformization procedure in obtaining the fundamental polygon.


\subsection{Whittaker's proposal}\label{secao_conjectura_Whittaker}

Whittaker in \cite{Whittaker1929} proposes the following result,

\begin{conjectura} \label{def_conjectura_Whittaker}
	Consider the hyperelliptic curve
	\begin{equation}\label{eq11}
	y^2=(z- e_1)(z- e_2)...(z- e_{2g+2})= f(z).
	\end{equation}
	
	The uniformization variable of (\ref{eq11}) is the quotient of two linearly independent solutions to the Fuchsian differential equation
	\begin{equation}\label{eq_2}
	\frac{d^2y}{dz^2}+\frac{3}{16}\left\{\sum_{r=1}^{2g+2} \frac{1}{(z-e_r)^2}+\frac{-(2g+2)z^{2g}+2gp_1z^{2g-1}+c_1z^{2g-2}+...+c_{2g-1}}{(z-e_1)(z-e_2)...(z-e_{2g+2})}\right\}y=0,
	\end{equation}
	where $p_1=\sum e_r$, and $c_1,...,c_{2g-1}$ are constants which depend on the roots of the hyperelliptic curve and are determined by the condition that the associated group is Fuchsian, \cite{Whittaker1898} and  \cite{Mursi1930}. Thus, (\ref{eq_2}) may be rewritten as
	\begin{equation*}
	\frac{d^2y}{dz^2}+\frac{3}{16}\left\{ \left[\frac{f'(z)}{f(z)}\right]^2-\frac{2g+2}{2g+1}\frac{f''(z)}{f(z)}\right\}y=0.
	\end{equation*}
\end{conjectura}

Whittaker's proposal holds when the solutions to the FDE lead to generators of a Fuchsian group whose associated fundamental region is the uniformization region of a hyperelliptic curve. The algebraic curves which are known to date and satisfy the Whittaker proposal are:
\begin{itemize}
	\item $y^2=z^{2g+1}+1$, \cite{Dhar1935};
	\item $y^2=z^{2g+2}-1$, \cite{Dalzell1930};
	\item $y^p=z^{2g+2}-1$, where $p$ divides $2g+2$, \cite{Dalzell1930};
	\item $y^2=z(z^4-1)$, \cite{burnside1893}.
\end{itemize}

Whittaker in \cite{Whittaker1929} considered the hyperelliptic curve $y^2=z^5+1$ as an instance for the determination of its uniformization region. Mursi, \cite{Mursi1930}, considered the case $y^2=z^7+1$. Dhar in \cite{Dhar1935}, proved the proposal for the case $y^2=z^{2g+1}+1$ and, Dalzell in \cite{Dalzell1930}, proved for the curves $y^2=z^{2g+2}-1$ and $y^p=z^{2g+2}-1$,where $p$ divides $2g+2$. The uniformization of the Bolza surface was carried out by Burnside in \cite{burnside1893}. However, Brezhnev in \cite{Brezhnev2009} uses a different approach when considering the case $y^2=z(z^4-1)$. This hyperelliptic curve is the affine model of the Bolza surface, and it is a compact Riemann surface of genus 2 with the highest possible order of the conformal automorphism group in genus 2.

\subsection{Classical procedure}\label{DC}

The classical procedure, as shown in \cite{Whittaker1929}, \cite{Mursi1930} and \cite{Dhar1935}, considers the hyperelliptic curve $y^2=z^{2g+1}+1$, $g \ge 2$. Hence, the linear differential equation is of the form $y''(z)+p(z)y'(z)+q(z)y(z)=0$, which may be reduced to $y''+Q(z)y=0$, \cite{kristensson2010}.

Once the FDE is specified, the procedure consists in realizing variable changes and algebraic manipulations to arrive at the hypergeometric differential equation. As mentioned in subsection \ref{sec_ed_complexa}, the HDE is given by
\[
z\left( z - 1\right) y''(z) + \left[ \left( \alpha + \beta + 1 \right) z - \gamma \right] y'(z) + \alpha \beta y(z) = 0,
\]
having three regular singular points at $z = 0$, $z = 1$ and $z = \infty$.

Under the geometric point of view, this is a hyperbolic triangle. There are two cases to consider leading to the same value taken by $g$. The first case is to consider the hyperbolic triangle having two generators, one being a parabolic transformation and the other one is an elliptic transformation. The second case is to consider the hyperbolic triangle having three generators, all three being elliptic transformations. Topologically, these regions are identified as a surface of genus $g=0$ and, therefore, a sphere. The second case is the one used in Whittaker's procedure. It is known that HDE has twenty-four solutions. If they are considered three-by-three, the solutions may be rewritten as a linear combination, see volume 1 of \cite{Bateman1953a}. The next step is to find the quotient of two linearly independent solutions of the HDE, leading to the generators (Mobius transformations) of the Fuchsian group.

\subsection{The uniformization procedure for the curve $y^2=z^{5}+1$} \label{sec_desenv}

Dhar in \cite{Dhar1935} has proved that the hyperelliptic curve $y^2 = z^{2g+1}+1 = f(z)$ satisfies Whittaker's proposal. Hence, the proper FDE, satisfying the condition that the group is Fuchsian, is given by
\begin{equation}\label{eqconjectura}
\frac{d^2y}{dz^2}+\frac{3}{16}\left[\left(\frac{f'(z)}{f(z)}\right)^2-\frac{2g+2}{2g+1} \frac{f''(z)}{f(z)}\right]y=0.
\end{equation}

Without loss of generality, consider the hyperelliptic curve $y^2=z^{5}+1=f(z)$. Although this was the curve considered by Whittaker, \cite{Whittaker1929}, we present next a short version of the steps to be followed in finding the uniformization region of the given curve. We refer the reader to \cite{Erika2019} and also to the Appendix where Whittaker's procedure is thoroughly revised to fill in important steps left out of in \cite{Whittaker1929} and \cite{Dhar1935}, a lengthy derivation, and to describe the whole procedure to obtain the generators of the Fuchsian group and the Fuchsian subgroup associated with the fundamental region which will uniformize the given hyperelliptic curve. We hope this effort may be helpful not only in shedding some light on this procedure as well as to highlight it as a promising alternative to be employed in the design of new digital communication systems.

\begin{example}
	Given $y^2=z^5+1$, a curve of genus $g=2$, it is known that this curve satisfies Whittaker's proposal, (\ref{eqconjectura}). Hence,
	\begin{equation}\label{eq111}
	\frac{d^2y}{dz^2}+\frac{3}{16}\left[\frac{5^2z^8}{(1+z^5)^2}-\frac{24 z^3}{1+z^5}\right]y=0
	\end{equation}
	
	From proper variable changes and algebraic manipulations, see Appendix, we arrive at the HDE given by
	
	\begin{equation}\label{eq222}
	x(1-x)u''(x)+\left[\frac{4}{5}-\frac{8}{5}x\right]u'(x)-\frac{2}{25}u(x)=0.
	\end{equation}
	Thus, the hypergeometric function associated with the HDE (\ref{eq222}) is given by
	$$F(\alpha, \beta; \gamma; x)=F\left(\frac{1}{5}, \frac{2}{5}; \frac{4}{5}; x\right).$$
	
	In order to simplify the notation, consider $a=\frac{1}{5}$. With this, $\alpha=a$, $\beta=2a$ and $\gamma=4a$. Therefore, the hyperelliptic curve $y^2=z^5+1$, whose associated FDE is given by (\ref{eq111}), leads to the HDE (\ref{eq222}) and the associated hypergeometric function is $$F(\alpha, \beta; \gamma; x)=F(a, 2a; 4a; x).$$
	
	Next, we find the solutions to the HDE (\ref{eq222}); see \cite{Sotomayor1979} for more detailed information. Note that the HDE has three singularities at $x=0$, $x=1$, and $x=\infty$, see subsection \ref{solucaohipergeometrica}.
	
	\begin{enumerate}
		\item[1-] The solutions at $x=0$ are
		\begin{itemize}
			\item  $w_{10}(x)=F(\alpha, \beta; \gamma; x)=F(a, 2a; 4a; x),$
			
			\item  $w_{20}(x)=x^a F(2a, 3a; 6a;x).$
		\end{itemize}
		
		\item[2-] The solutions at $x=1$ are
		\begin{itemize}
			\item  $w_{11}(x)=F(a, 2a; 4a; 1-x),$
			
			\item  $w_{21}(x)=(1-x)^a F(2a, 3a; 6a;1-x).$
		\end{itemize}
		
		\item[3-] The solutions at $x=\infty$ are
		\begin{itemize}
			\item  $w_{1\infty}(x)=x^{-a}F(a, 2a; 4a; x^{-1}),$
			
			\item  $w_{2\infty}(x)=x^{-2a} F(2a, 3a; 6a;x^{-1}).$
		\end{itemize}
	\end{enumerate}
	
	Relating the solutions $w_{10}(x)$ with $w_{1\infty}(x)$ and $w_{2\infty}(x)$, and $w_{20}(x)$ with $w_{1\infty}(x)$ and $w_{2\infty}(x)$, and using $t=w_{20}(x)/w_{10}(x)$ as the uniformizing variable, with $t$ denoting $w_{20}(x)/w_{10}(x)$ at $i\infty$ and $t'$ denoting $w_{20}(x)/w_{10}(x)$ at $-i\infty$, after variable changes and algebraic manipulations, see Appendix, we end up with the uniformizing variable.
	
	\begin{equation}\label{eqyy}
	t'_1=\frac{[2(\cos \pi a) \exp(-3\pi ai)]t_1 -i(\cos2\pi a +\cos \pi a)/\sin \pi a}{(2i\sin \pi a)t_1 + 2(\cos \pi a) exp(3\pi ai)}.
	\end{equation}
	
	Relating the solutions $w_{11}(x)$, with $w_{1\infty}(x)$ and $w_{2\infty}(x)$, and $w_{21}(x)$ with $w_{1\infty}(x)$ and $w_{2\infty}(x)$, similarly as in the previous procedure, the quotient of the two solutions undergoes the following transformation as $x$ passes from $i\infty$ to $-i\infty$, rounding $x=0$,
	\begin{equation}\label{eqrr}
	t'_1 = \{\exp(2\pi ai)\}t_1.
	\end{equation}
	Substituting the following transformations
	\[
	t_1 =it_2 (\cot \pi a) \exp(2\pi ai),
	\]
	and
	\[
	t_2 = \sqrt{\frac{(\cos 2\pi a + \cos \pi a)}{\cos ^2 \pi a}}t_3,
	\]
	in (\ref{eqrr}) and (\ref{eqyy}), we end up
	\[
	t'_3=\frac{\{\exp(-\pi ai)\}\sqrt{\frac{(\cos 2\pi a + \cos \pi a)}{\cos ^2 \pi a}}t_3 - 1}{\{\exp(-\pi ai)\}t_3 -\sqrt{\frac{(\cos 2\pi a + \cos \pi a)}{\cos ^2 \pi a}}^{-1}},
	\]
	and
	\[
	t'_3 = \{\exp(2\pi ai)\}t_1.
	\]
	
	Thus, knowing the transformations $t'_1$, and $t'_3$ and its inverse, it follows that $t'_1 \circ (t'_3)^{-1}$ is the resulting transformation of the quotient of the two solutions of (\ref{eq222}) when $z$ passes successively by the two circuits, namely, from the infinite to the singularity $z_1=e^{a\pi i}$, rounding it, and returning to infinite, and from the infinite to the singularity $z_2=e^{3a\pi i}$, rounding it, and returning to infinite. Therefore, the Mobius transformations are given by
	
	\[
	S_j(t)=\frac{\left(2\cos (a\pi) -1\right)^{-1/2} t-e^{\frac{1}{2}(4k+1)a\pi i}}{e^{-\frac{1}{2}(4k+1)a\pi i}t-\left(2\cos (a\pi) -1\right)^{-1/2}},
	\]
	or in the matrix form by
	\[
	S_j=\left(
	\begin{array}{ll}
	\left(2\cos (a\pi) -1\right)^{-1/2} & -e^{\frac{1}{2}(4k+1)a\pi i} \\
	&    \\
	e^{-\frac{1}{2}(4k+1)a\pi i} & -\left(2\cos (a\pi) -1\right)^{-1/2} \\
	\end{array}
	\right),
	\]
	where $a=1/5$, $j=1,\ldots, 5$ and $k=0,1, 2, 3,4$ are the generators of the Fuchsian group $\Gamma _0$.
	
	Fig. \ref{fig_wittaker} shows the polygon in $\mathbb{D}^2$ resulting from the Whittaker procedure.

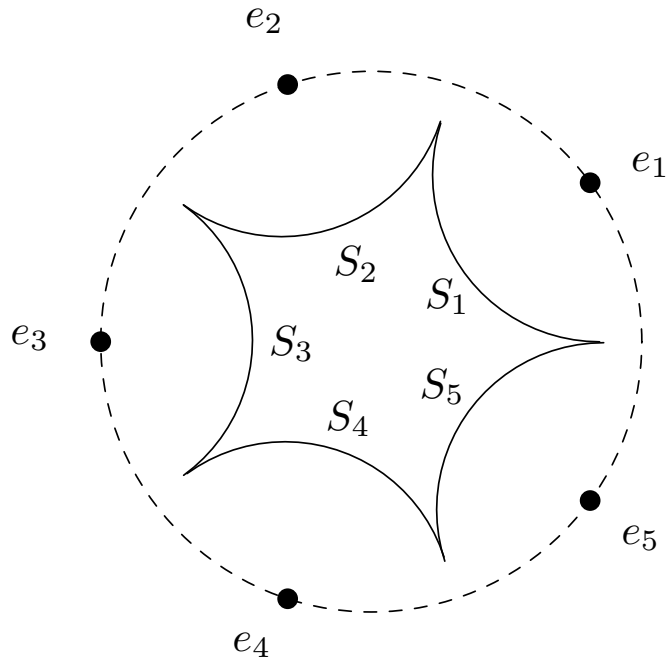
\begin{figure}[h!]
\centering
\resizebox{0.55\linewidth}{!}{\begin{tikzpicture}[line cap=round,line join=round,>=triangle 45,x=1.0cm,y=1.0cm]
			\begin{scope}[xshift=3cm, yshift=1cm, scale=2.25]
			\draw [dashed] (0.,0.) circle (1.cm);
			\begin{scope}[scale=0.85]
			\draw [shift={(0.38196596008601663,1.1755703764586758)}, rotate=36, xshift=-0.62cm, yshift=0.46cm]  plot[domain=3.455751834461369:5.340707524337482,variable=\t]({1.*0.7265423550317994*cos(\t r)+0.*0.7265423550317994*sin(\t r)},{0.*0.7265423550317994*cos(\t r)+1.*0.7265423550317994*sin(\t r)});
			\draw [shift={(-1.,0.7265425455885678)},  rotate=36, xshift=-0.62cm, yshift=-0.44cm]  plot[domain=-1.5707963267948966:0.3141591808715763,variable=\t]({1.*0.7265425455885678*cos(\t r)+0.*0.7265425455885678*sin(\t r)},{0.*0.7265425455885678*cos(\t r)+1.*0.7265425455885678*sin(\t r)});
			\draw [shift={(-1.,-0.7265425455885678)},  rotate=36, xshift=0.25cm, yshift=-0.72cm]  plot[domain=-0.31415918087157646:1.5707963267948966,variable=\t]({1.*0.726542607292283*cos(\t r)+0.*0.726542607292283*sin(\t r)},{0.*0.726542607292283*cos(\t r)+1.*0.726542607292283*sin(\t r)});
			\draw [shift={(0.38196596008601663,-1.1755703764586758)}, rotate=36, xshift=0.77cm, yshift=-0.01cm]  plot[domain=0.942477782842104:2.827433472718217,variable=\t]({1.*0.7265424412558247*cos(\t r)+0.*0.7265424412558247*sin(\t r)},{0.*0.7265424412558247*cos(\t r)+1.*0.7265424412558247*sin(\t r)});
			\draw [shift={(1.236067857017757,0.)},rotate=36, xshift=0.23cm, yshift=0.73cm]  plot[domain=2.199114870747689:4.084070436431897,variable=\t]({1.*0.7265424703545018*cos(\t r)+0.*0.7265424703545018*sin(\t r)},{0.*0.7265424703545018*cos(\t r)+1.*0.7265424703545018*sin(\t r)});
			\end{scope}
			\begin{scriptsize}
			\draw [fill=black] (0.8090169,0.5877852) circle (1.pt);
			\draw[color=black] (0.9135102479476468,0.6529126077136448) node[right] {\small $e_1$};
			\draw [fill=black] (-0.3090169,0.9510565) circle (1pt);
			\draw[color=black] (-0.395013723446097,1.094279820286931) node[above] {\small $e_2$};
			\draw [fill=black] (-1.,0.) circle (1.pt);
			\draw[color=black] (-1.1427417070996648,0.011807939718366262) node[left] {\small $e_3$};
			\draw [fill=black] (-0.3090169,-0.9510565) circle (1.pt);
			\draw[color=black] (-0.3223179472575557,-1.0294753555068814) node[below left] {\small $e_4$};
			\draw [fill=black] (0.8090169,-0.5877852) circle (1.pt);
			\draw [color=black] (0.8771623598533761,-0.6348411419119433) node[below right] {\small $e_5$};
			\draw[color=black] (0.15135102479476468,0.16529126077136448) node[right] {\small $S_1$};
			\draw[color=black] (-0.185135102479476468,0.286529126077136448) node[right] {\small $S_2$};
			\draw[color=black] (-0.425135102479476468,-0.006529126077136448) node[right] {\small $S_3$};
			\draw[color=black] (-0.215135102479476468,-0.286529126077136448) node[right] {\small $S_4$};
			\draw[color=black] (0.13215135102479476468,-0.16529126077136448) node[right] {\small $S_5$};
			\end{scriptsize}
			\end{scope}
			\end{tikzpicture}}
\caption{Fundamental region: regular pentagon}
\label{fig_wittaker}
\end{figure}
\end{example}

Note in Fig. \ref{fig_wittaker} that the vertices originating from the Mobius transformations $S_j$, with $j=1,2,3,4,5$, are inside $\mathbb{D}^2$. The Fuchsian group generated by $S_j$ (elliptic transformations), with $j=1,2,3,4,5$, is denoted by $\Gamma _0$ since this pentagon is the fundamental polygon of a surface with $g=0$. The uniformizing region for the curve $y^2 = z^5 +1$ is the juxtaposition of two of these polygons giving rise to a polygon with eight sides and genus $g=2$. The corresponding generators are given by $S_2S_1$, $S_3S_1$, $S_4S_1$, and $S_5S_1$ leading to the Fuchsian subgroup $\Gamma_8$. The difference between this procedure and the one to be considered in the next section is that the vertices arising from the Mobius transformations are at the boundary of $\mathbb{D}^2$ denoted by $\partial \mathbb{D}^2$.

\section{Procedure to Find the Uniformization Region Whose Vertices are at the Boundary of $\mathbb{D}^2$} \label{cap_Primeiros_Resultados}

The objective of this section is to determine the generators of the Fuchsian group associated with the polygon whose vertices are at the boundary of $\mathbb{D}^2$, when the hyperelliptic curve $y^2 = z^{(2g+1)} \pm 1$ is considered. Since the curve $y^2 = z^{(2g+2)} \pm 1$ has the same number of branch points as the curve $y^2 = z^{(2g+1)} \pm 1$ it follows that there is no need to consider it.

In order to find the uniformization region, we make use of Whittaker's procedure to obtain the generators of the Fuchsian group. Next, by use of hyperbolic isometries, the vertices of this region are allocated in the boundary of the Poincaré disk. Since the allocated vertices are similar, up to rotation, to the roots of the hyperelliptic curve, it follows that the procedure then makes use of the roots of the given hyperelliptic curve ($y^2 = z^{(2g+1)}\pm 1$) as the vertices to be used in the determination of the generators of the Fuchsian group, $\Gamma_{0}$. Note that the vertices at the boundary of the disk reduces the computational complexity substantially in finding the generators of the Fuchsian group, and so rather than solving the FDE, we use the strategy that the fundamental polygon has its vertices at the boundary of the Poincaré disk. This strategy is twofold: it is relatively easy to determine the Mobius transformations, and the signal decision region has the most significant area leading to the least error probability.

Note that determining the uniformization region is equivalent to knowing its number of sides, $p$. Thus, a relation between the degree of the hyperelliptic curve and the number of sides of the polygon associated with the fundamental region may be established.

Therefore, the procedure begins by determining the vertices of the polygon in $\mathbb{D}^2$, which are the roots of the hyperelliptic curve. Knowing these roots, the sides of this regular hyperbolic polygon consist of geodesics connecting pair of subsequent vertices. Thus, given two distinct points, $z_1$ and $z_2$, there exists a unique geodesic in $\mathbb{D}^2$ connecting $z_1$ to $z_2$, \cite{Anderson2008}. Furthermore, the geodesics are characterized by use of the cross-ratio which are also elliptic Mobius transformations given by $$S_j(z)=\frac{az+b}{cz+d} \qquad \longleftrightarrow \qquad S_j = \left(
\begin{array}{cc}
a & b \\
c & d \\
\end{array}
\right),
$$
with $ad-bc=1$ and $Tr(S_j)= a+d < 2$, where $j=1,\ldots , n$ and $n$ is the degree of the hyperelliptic curve.

Fixing one of these elliptic transformations, for instance, $S_k$, and multiplying it by the remaining elliptic transformations, $S_j$ for every $j \neq k$, lead to hyperbolic transformations. This new polygon has $2(n-1)$ sides, where $n$ denotes the number of sides of the polygon arising from the roots of the hyperelliptic curve, and this region is the fundamental region which uniformizes the hyperelliptic curve.

Since the elliptic transformations satisfy the condition that the determinant is 1, then $S_j^2=I$, where $I$ denotes the identity matrix. Hence, the Fuchsian group is obtained whose generators are the elliptic transformations as the result of the quotients of the linearly independent solutions of the hypergeometric differential equation. Next, by use of the Scilab software, the generators of the Fuchsian subgroup $\Gamma_{8} = \langle S_kS_j\rangle $ are determined. Recall that the subgroup $\Gamma_{8}$ identifies the fundamental region of the compact surface, which uniformizes the hyperelliptic curve.

As a consequence of the procedure where the singularities or vertices of the polygon are at the boundary of $\mathbb{D}^2$, we present next the steps of the algorithm to find the Fuchsian subgroup.

\begin{center}
	\textbf{Algorithm to Obtain the Fuchsian Subgroup in $\mathbb{D}^2$} \label{secao_Algoritmo}
\end{center}

\begin{itemize}
	\item[Step 1-] Select a set of $2g+1$ symmetric points in $\mathbb{D}^{2}$ as the roots of the hyperelliptic curve;
	
	\item[Step 2-] For each two subsequent vertices of the hyperbolic polygon, determine the elliptic transformation,
	$$S_j(z)=\frac{az+b}{cz+d} \qquad \longleftrightarrow \qquad  S_j = \left(                                                                                                                                       \begin{array}{cc}                                                                                                                                         a & b \\
	c & d \\                                                                                                                                      \end{array}
	\right)
	$$ with $ad-bc= 1$, $|tr(S_j)|<2$ and $j=1, \ldots , (2g+1)$; \label{cond1_disco}
	
	\item[Step 3-] The generators of the Fuchsian group $\Gamma _{0}$ are specified by $S_j$, $j=1,\ldots, (2g+1)$;
	
	\item[Step 4-] Fix one of the elliptic transformations, for instance $S_k$, and multiply it by all the remaining elliptic transformations $S_kS_j$, for every $j \neq k$, and verify if every $S_kS_j$ is hyperbolic. The resulting set of such transformations consists of the generators of the Fuchsian subgroup $\Gamma_{4g}$;
	
	\item[Step 5-] Knowing $g$ and $p$, either select $q$ from the inequality $(p-2)(q-2)>4$ or find $q$ by using the Euler characteristic $\chi (\Gamma _{p})=p/q - p/2 + 1 = 2 - 2g$.
\end{itemize}

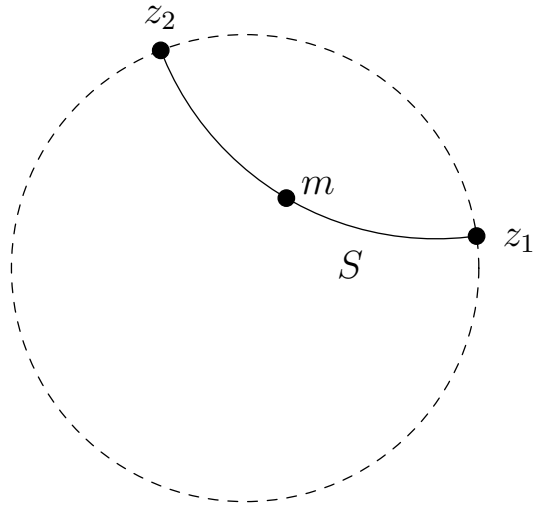
\begin{figure}[h!]
	\centering
	\resizebox{0.45\linewidth}{!}{
		\begin{tikzpicture}[line cap=round,line join=round,>=triangle 45,x=1.0cm,y=1.0cm]
		\begin{scope}[xshift=3cm, yshift=1cm, scale=2.5]
		\draw [dashed] (0.,0.) circle (1.cm);
		\draw [shift={(0.8171533085383949,1.388833174719995)}]  plot[domain=3.5110798443628326:4.850043920020342,variable=\t]({1.*1.2635652404439057*cos(\t r)+0.*1.2635652404439057*sin(\t r)},{0.*1.2635652404439057*cos(\t r)+1.*1.2635652404439057*sin(\t r)});
		\begin{scriptsize}
		\draw [fill=black] (-0.3611372783729415,0.9325126627394315) circle (1pt);
		\draw[color=black] (-0.3627775809202965,1.0750521064272136) node {{\large $z_2$}};
		\draw [fill=black] (0.9905405101897948,0.1372206167925979) circle (1pt);
		\draw[color=black] (1.0602000078048197,0.119588760893462753) node[right] {{\large $z_1$}};
		\draw[color=black] (0.3509771013223972,0.11658153312988392) node[below right] {{\large$S$}};
		\draw[fill=black]   (0.17638809177542566,0.29978907375586483) circle (1pt);
		\draw (0.31210492052753994,0.35240208340602696085) node {{\large$m$}};
		\end{scriptsize}
		\end{scope}
		\end{tikzpicture}
	}
	\caption{Geodesic between $z_1$ and $z_2$ in $\mathbb{D}^2$}
	\label{fig_geodesica_geral_disco}
\end{figure}

Given $z_1$ and $z_2$, two consecutive vertices of the regular hyperbolic polygon and the midpoint $m$, by use of the cross-ratio the geodesic connecting these two vertices is obtained and it is perpendicular to the boundary of $\mathbb{D}^2$, see Fig. \ref{fig_geodesica_geral_disco}. This geodesic is given by
$$w=\frac{[z_1(m-z_2)^2 - z_2(m-z_1)^2]z+ [z_2^2(m-z_1)^2- z_1^2(m-z_2)^2]}{[(m-z_2)^2 - (m-z_1)^2]z+ [z_2(m-z_1)^2 - z_1(m-z_2)^2]}.$$
Since it is an elliptic transformation, the vertex $z_1$ is taken to vertex $z_2$, and vice-versa, and the midpoint $m$ is a fixed point.

The transformation $w$ has determinant not equal to one. Therefore, normalization is necessary. This normalization is achieved if we multiply $w$ by $1/det(w)$. Hence, the elliptic transformation $S$, acting on the side of the regular hyperbolic polygon in $\mathbb{D}^2$ is given by

$$S=\frac{\frac{[z_1(m-z_2)^2 - z_2(m-z_1)^2]}{\det (w)}z+ \frac{[z_2^2(m-z_1)^2- z_1^2(m-z_2)^2]}{\det (w)}}{\frac{[(m-z_2)^2 - (m-z_1)^2]}{\det (w)}z+ \frac{[z_2(m-z_1)^2 - z_1(m-z_2)^2]}{\det (w)}}.
$$

\subsection{Example} \label{sec_aplicacao_em_exemplos}

In this subsection, we make use of the previous algorithm when the curve is $y^2=z^{2g+1}-1$, for $g=2$.

\begin{example}\label{example_algoritmo_disco_z5menos1}
	Given the curve $y^2=z^5-1$, of genus 2, its roots are $c_1 = 0.3090+0.9511i$, $ c_2 = -0.8090+0.5878i$, $c_3 =  -0.8090-0.5878i$, $c_4 = 0.3090-0.9511i$ and $c_5 = 1$. These roots are at the boundary of $\mathbb{D}^2$, see Fig. \ref{fig_pto_medio_geodesica_z5menos1_disco}, and the polygon is regular.
	
	\begin{figure}[h!]
		\centering
		\resizebox{0.5\linewidth}{!}{
			\begin{tikzpicture}[line cap=round,line join=round,>=triangle 45,x=1.0cm,y=1.0cm]
			\begin{scope}[xshift=3cm, yshift=1cm, scale=2.5]
			\draw [dashed] (0.,0.) circle (1.cm);
			\draw [shift={(-0.38196601125013663,1.17557050458488)}]  plot[domain=4.0840704496667435:5.969026041820747,variable=\t]({1.*0.7265425280052393*cos(\t r)+0.*0.7265425280052393*sin(\t r)},{0.*0.7265425280052393*cos(\t r)+1.*0.7265425280052393*sin(\t r)});
			\draw [shift={(-1.2360679774998622,0.)}]  plot[domain=-0.942477796076826:0.9424777960768262,variable=\t]({1.*0.726542528005453*cos(\t r)+0.*0.726542528005453*sin(\t r)},{0.*0.726542528005453*cos(\t r)+1.*0.726542528005453*sin(\t r)});
			\draw [shift={(-0.38196601125013663,-1.17557050458488)}]  plot[domain=0.3141592653588387:2.1991148575128427,variable=\t]({1.*0.7265425280053109*cos(\t r)+0.*0.7265425280053109*sin(\t r)},{0.*0.7265425280053109*cos(\t r)+1.*0.7265425280053109*sin(\t r)});
			\draw [shift={(1.,-0.7265425280053753)}]  plot[domain=1.5707963267948966:3.455751918948794,variable=\t]({1.*0.7265425280053753*cos(\t r)+0.*0.7265425280053753*sin(\t r)},{0.*0.7265425280053753*cos(\t r)+1.*0.7265425280053753*sin(\t r)});
			\draw [shift={(1.,0.7265425280053753)}]  plot[domain=2.827433388230792:4.71238898038469,variable=\t]({1.*0.726542528005416*cos(\t r)+0.*0.726542528005416*sin(\t r)},{0.*0.726542528005416*cos(\t r)+1.*0.726542528005416*sin(\t r)});
			\begin{scriptsize}
			\draw [fill=black] (0.3090169943749,0.9510565162952) circle (1pt);
			\draw[color=black] (0.35020269789402575,1.042433349201057) node {{\large $c_1$}};
			\draw [fill=black] (-0.8090169943749,0.5877852522925) circle (1pt);
			\draw[color=black] (-0.9153052029813895,0.6839447629462484) node {{\large $c_2$}};
			\draw [fill=black] (-0.8090169943749,-0.5877852522925) circle (1pt);
			\draw[color=black] (-0.9412200405419783,-0.65686057191488652) node {{\small $c_3$}};
			\draw [fill=black] (0.3090169943749,-0.9510565162952) circle (1pt);
			\draw[color=black] (0.357611753545461446,-1.1005196784920078) node {{\small $c_4$}};
			\draw [fill=black] (1.,0.) circle (1pt);
			\draw[color=black] (1.15088775568370804,-0.002798432409348302) node {{\small $c_5$}};
			\draw[color=black] (-0.3970084517696153,0.3384075198289666372) node {{\small $T_1$}};
			\draw[color=black] (-0.435006678020119703,-0.22739369126778242) node {{\small $T_2$}};
			\draw[color=black] (0.09537346188157012,-0.504822229272802366) node {{\small $T_3$}};
			\draw[color=black] (0.447113860651010636,-0.06711757200277973) node {{\small $T_4$}};
			\draw[color=black] (0.20292667892779451,0.48958348124183426) node {{\small $T_5$}};
			\draw [fill=black] (-0.15745202296032845,0.4845874989599603) circle (1.0pt);
			\draw[color=black] (-0.21560458884549438,0.596234768086382084) node {{\small $m_1$}};
			\draw [fill=black] (-0.5095254494944091,0.) circle (1pt);
			\draw[color=black] (-0.67259228502214823,0.010518005742643735) node {{\small $m_2$}};
			\draw [fill=black] (-0.15745202296030636,-0.4845874989598923) circle (1pt);
			\draw[color=black] (-0.152415194264060831,-0.625383717419948452) node {{\small $m_3$}};
			\draw [fill=black] (0.4122147477075104,-0.29949154488053104) circle (1.0pt);
			\draw[color=black] (0.584884151648838322,-0.3213776483136411) node {{\small $m_4$}};
			\draw [fill=black] (0.4122147477074775,0.29949154488050717) circle (1.0pt);
			\draw[color=black] (0.5814625003273232434,0.328160499140604864) node {{\small $m_{5}$}};
			\end{scriptsize}
			\end{scope}
			\end{tikzpicture}
		}
		\caption{Polygon arising from the roots of $z^5-1=0$ in $\mathbb{D}^2$}
		\label{fig_pto_medio_geodesica_z5menos1_disco}
	\end{figure}
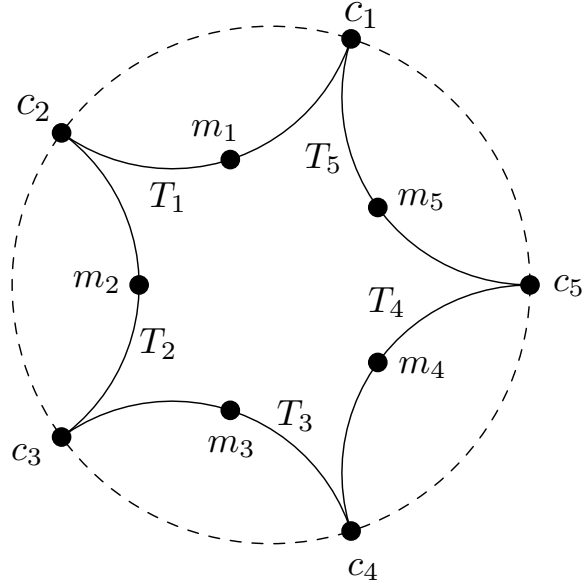
	
	By using the cross-ratio, the geodesic associated with each side of the regular hyperbolic polygon and its midpoint are determined. From this, the corresponding elliptic transformations are obtained. The midpoint of each geodesic is given by $m_1=-0.1575+0.4846i$, $m_2 = -0.5095$, $m_3=-0.1575-0.4846i$, $m_4 = 0.4122-0.2995i$ and $m_5=0.4122+0.2995i$, as shown in Fig. \ref{fig_pto_medio_geodesica_z5menos1_disco}. The elliptic transformation acting on the side containing $c_1$ and $c_2$, for instance, comes from the use of the cross-ratio given by
	$$w=\frac{[c_1(m_1-c_2)^2 -c_2(m_1-c_1)^2]z+ [c_2^2(m_1-c_1)^2 - c_1^2(m_1-c_2)^2]}{[(m_1-c_2)^2-(m_1-c_1)^2]z+ [c_2(m_1-c_1)^2 -c_1(m_1-c_2)^2]},$$
	where $m_1$ is the midpoint of this side. Since $det(w)\neq 1$, after normalization, the elliptic transformation $T_1$ is given by
	$$T_1=\frac{1.7013i  z+(1.30902 +0.425325i)}{(1.30902 -0.425325i)z -1.7013i}.$$
	
	Similarly, the remaining elliptic transformations are given by
	$$T_2=\frac{-1.7013iz-1.3763i}{1.3764iz+1.7013i} \qquad T_3=\frac{1.7013iz+(-1.3090+0.4253i)}{(-1.3090-0.4253i)z-1.7013i},$$
	$$T_4=\frac{-1.7013iz+(0.8090+1.1135i)}{(0.8090-1.1135i)z+1.7013i} \qquad T_5=\frac{-1.7013i z+(-0.8090+1.1135i)}{(-0.8090-1.1135i)z+1.7013i}.$$
	
	Note that $tr(T_j)=0$ and $det(T_j)= 1$, for $j=1, \ldots, 5$. In addition to being elliptic transformations they are also the generators $T_1$, $T_2$, $T_3$, $T_4$ and $T_5$ of the Fuchsian group $\Gamma_{0}$ associated with the hyperelliptic curve.
	
	Note that, from Fig. \ref{fig_pto_medio_geodesica_z5menos1_disco}, the polygon in $\mathbb{D}^2$ arising from the roots of $y^2=z^5-1$ is regular. In this way, we may select $T_1$ as the fixed elliptic transformation and multiply it by the remaining elliptic transformations. Thus,
	$$T_1T_2=\frac{(2.3090+1.8017i)z+(1.6180+2.2270i)}{(1.6180-2.2270i)z+(2.3090-1.8017i)},$$
	$$T_1T_3=\frac{(-4.4271-1.1135i)z-4.4541i}{4.4541iz+(-4.4271+1.1135i)},$$
	$$T_1T_4=\frac{(4.4271-1.1135i)z+(-2.6180+3.6034i)}{(-2.6180-3.6034i)z+(4.4271+1.1135i)},$$
	and
	$$T_1T_5=\frac{(2.3090-1.8017i)z+(-2.6180+0.85066i)}{(-2.6180-0.85066i)z+(2.3090+1.8017i)}.$$
	
	Knowing the previous results, we have to verify if the transformations $T_1T_j$, where $j=2,3,4,5$, are hyperbolic, that is, the trace is a real number whose absolute value is greater than 2. Hence, $|Tr(T_1T_2)|=4.6180$, $|Tr(T_1T_3)|=8.8541$, $|Tr(T_1T_4)|=8.8541$ and $|Tr(T_1T_5)|=4.6180$. Therefore, the generators of the Fuchsian subgroup $\Gamma_{8}$, forming the fundamental region, are hyperbolic transformations $T_1T_2$, $T_1T_3$, $T_1T_4$ e $T_1T_5$, that is, $\Gamma_{8}=\langle T_1T_2, T_1T_3, T_1T_4, T_1T_5\rangle$.
	
	Fig. \ref{fig_reg_fundamental_z5menos1_disco} shows the fundamental region which uniformizes the given hyperelliptic curve.
	
	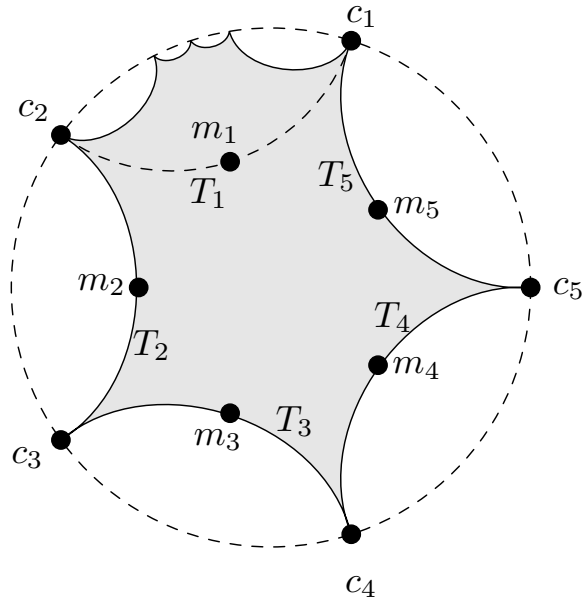
\begin{figure}[h!]
		\centering
		\resizebox{0.5\linewidth}{!}{
			\begin{tikzpicture}[line cap=round,line join=round,>=triangle 45,x=1.0cm,y=1.0cm]
			\begin{scope}[xshift=3cm, yshift=1cm, scale=2.5]
			\draw [dashed] (0.,0.) circle (1.cm);
			\draw[fill=gray!20!]    (0.3090169943749,0.9510565162952)
			to [bend left=70]  (-0.16035745659102058,0.9870590084263751)
			to [bend left=70]  (-0.30901699437499636,0.9510565162951377)
			to [bend left=70]  (-0.45044682073865366,0.8928032603471154)
			to [bend left=70]  (-0.8090169943749,0.5877852522925)
			to [bend left=58] (-0.8090169943749,-0.5877852522925)
			to [bend left=58]   (0.3090169943749,-0.9510565162952)
			to [bend left=58] (1.,0.)
			to [bend left=58]  (0.3090169943749,0.9510565162952);
			\draw [shift={(-0.38196601125013663,1.17557050458488)},dashed]  plot[domain=4.0840704496667435:5.969026041820747,variable=\t]({1.*0.7265425280052393*cos(\t r)+0.*0.7265425280052393*sin(\t r)},{0.*0.7265425280052393*cos(\t r)+1.*0.7265425280052393*sin(\t r)});
			\begin{scriptsize}
			\draw [fill=black] (0.3090169943749,0.9510565162952) circle (1pt);
			\draw[color=black] (0.3513001575614501,1.0423506206707471) node {{\small $c_1$}};
			\draw [fill=black] (-0.8090169943749,0.5877852522925) circle (1pt);
			\draw[color=black] (-0.9124643730122763,0.6806671691597492) node {{\small $c_2$}};
			\draw [fill=black] (-0.8090169943749,-0.5877852522925) circle (1pt);
			\draw[color=black] (-0.9379949695895233,-0.65703320631253497) node {{\small $c_3$}};
			\draw [fill=black] (0.3090169943749,-0.9510565162952) circle (1pt);
			\draw[color=black] (0.34768307541386971,-1.1519958420060794649) node {{\small $c_4$}};
			\draw [fill=black] (1.,0.) circle (1pt);
			\draw[color=black] (1.147432358872072,-0.004403838996376391) node {{\small $c_5$}};
			\draw[color=black] (-0.2393342242608254,0.3640834080566911546) node {{\small $T_1$}};
			\draw[color=black] (-0.459971972834678324,-0.2214139099029752) node {{\small $T_2$}};
			\draw[color=black] (0.09173909235943897,-0.5148097497510498544) node {{\small $T_3$}};
			\draw[color=black] (0.4704429415886028,-0.1129140378554587) node {{\small $T_4$}};
			\draw[color=black] (0.253215737353429747,0.42891876948303973) node {{\small $T_5$}};
			\draw [fill=black] (-0.15745202296032845,0.4845874989599603) circle (1.0pt);
			\draw[color=black] (-0.21462806656752503,0.59721095777146173) node {{\small $m_1$}};
			\draw [fill=black] (-0.5095254494944091,0.) circle (1.0pt);
			\draw[color=black] (-0.6523117611803477,0.010197364674215242) node {{\small $m_2$}};
			\draw [fill=black] (-0.15745202296030636,-0.4845874989598923) circle (1.0pt);
			\draw[color=black] (-0.204015866314477202,-0.586362912676890205) node {{\small $m_3$}};
			\draw [fill=black] (0.4122147477075104,-0.29949154488053104) circle (1.0pt);
			\draw[color=black] (0.56148746333930676744,-0.31077099792333934) node {{\small $m_4$}};
			\draw [fill=black] (0.4122147477074775,0.29949154488050717) circle (1.0pt);
			\draw[color=black] (0.5646193274272952045,0.30855510966232736) node {{\small $m_5$}};
			\end{scriptsize}
			\end{scope}
			\end{tikzpicture}
		}
		\caption{Fundamental region associated with $y^2=z^5-1$ in $\mathbb{D}^2$}
		\label{fig_reg_fundamental_z5menos1_disco}
	\end{figure}
\end{example}

This example is just a glimpse of the many cases considered in \cite{Erika2019}.

Lastly, by exhibiting the fundamental region which uniformizes the given hyperelliptic curve, it becomes possible to tile the Poincaré disk. Nevertheless, the Fuchsian subgroup may not be sufficient to determine which tessellation must be used. It is necessary to take into consideration the degree of the hyperelliptic curve. However, once the tessellation is known, the action of a proper Fuchsian subgroup on the uniformizing region leads to the desired signal constellation.

\subsection{Relationship between the hyperelliptic curve and the associated tessellation}\label{secao_relacao_curva_tesselacao}

The objective is to obtain a relation between the hyperelliptic curve degree and the parameters of the regular tessellation. Among the infinite tessellations in the hyperbolic plane, we call attention to the following ones: the self-dual tessellation $\{4g,4g\}$, $\{4g+2,2g+1\}$, and the densest tessellation $\{12g-6,3\}$, where $g=0,1, 2, \ldots, $ is the genus of the associated surface. These tessellations have rich algebraic and geometric structures that are beginning to be explored for applications in classical and quantum coding theory.

Retaking the example of subsection \ref{sec_aplicacao_em_exemplos}, we notice that, given a hyperelliptic curve of degree $n$, $n\in \mathbb{N}$ and $n\geq 3$, its roots lead to an $n$ side polygon. The juxtaposition of two such polygons leads to a fundamental polygon with $p=2n-2$ sides. Hence, $q$ can take any value satisfying the inequality $(p-2)(q-2)>4$, see Theorem 7.3.1 in  \cite{Walkden}.

In particular, our interest is in the hyperelliptic curves of the form $y^2=z^{2g+1}\pm 1$ since it is from these curves that arithmetic Fuchsian groups may be determined leading to complete labeling of the corresponding hyperbolic lattices with applications in digital transmission as well as in topological quantum codes. Thus, for the hyperelliptic curve of degree $2g+1$, it follows that $p=2(2g+1)-2=4g+2-2=4g$. Then, the tessellation is $\{4g,q\}$, where $q$ has to satisfy the condition $(p-2)(q-2)>4$. Similarly, given the hyperelliptic curve of degree $2g+2$, then $p=2(2g+2)-2=4g+4-2=4g+2$. Hence, the tessellation is $\{4g+2,q\}$, where $q$ has to satisfy the condition $(p-2)(q-2)>4$. In special, we consider the hyperelliptic curve of degree $6g-2$ and so, $p=2(6g-2)-2=12g-4-2=12g-6$. Then, the tessellation is $\{12g-6,q\}$, where $q$ has to satisfy the condition $(p-2)(q-2)>4$. Hence, we have the following result.

\begin{proposition}\label{prop_tesselacao2}
	Consider the hyperelliptic curve of degree $n$ with $n$ distinct roots in $\mathbb{D}^2$ forming a regular hyperbolic polygon with $n$ sides. From the juxtaposition of two of these polygons leads to a polygon with $p=2(n-1)$ sides. If the degree of the curve is
	\begin{enumerate}
		\item $2g+1$ then the associated tessellation is $\{4g,q\}$, where $q$ satisfies $(4g-2)(q-2)>4$;
		\item $2g+2$ then the associated tessellation is $\{4g+2,q\}$, where $q$ satisfies $4g(q-2)>4$;
		\item $6g-2$ then the associated tessellation is $\{12g-6,q\}$, where $q$ satisfies $(12g-8)(q-2)>4$;
	\end{enumerate}
	where $g$ is the genus of the associated surface.
\end{proposition}

Note that Proposition \ref{prop_tesselacao2} establishes all the possible tessellations of $\mathbb{D}^2$ for a given number of sides $p$. However, the following result shows that for a given compact surface, there is a specific value that $q$ may take. This comes from the Euler characteristic given by $\chi (\Gamma _{p})=p/q -p/2 +1 = 2-2g$.

\begin{proposition}\label{prop_tessel2}
	Let $g$ be a given value of the genus of a compact surface, with $g\ge 2$. Consider the hyperelliptic curve of degree $n$ with $n$ distinct roots in $\mathbb{D}^2$ forming a regular hyperbolic polygon with $n$ sides. The juxtaposition of two of these polygons leads to a polygon with $p=2(n-1)$ sides.
	\begin{enumerate}
		\item If $n=2g+1$, then from the Euler characteristic $q=4g$, leading to a self-dual tessellation $\{4g,4g\}$, satisfying $(p-2)(q-2)=(4g-2)(4g-2)>4$;
		\item If $n=2g+2$, then from the Euler characteristic $q=2g+1$, leading to the tessellation $\{4g+2,2g+1\}$, satisfying $(p-2)(q-2)=(4g)(2g-1)>4$;
		\item If $n=6g-2$, then from the Euler characteristic $q=3$, leading to the densest tessellation $\{12g-6,3\}$, where $q$ satisfies $(p-2)(q-2)=(12g-8)(1)>4$.
	\end{enumerate}
\end{proposition}

\section{Conclusions}\label{conclusoes}

We have presented in this paper a new approach to the construction of hyperbolic signal sets matched to groups by use of Whittaker's proposal for the uniformization of hyperelliptic functions via Fuchsian differential equations (FDEs). The aim was to provide a systematic approach to the design of new digital communication systems. This approach consisted of the four steps described next:
\begin{itemize}
	\item[1) ] Let $C$ be a DMC and $K_{m,n}$ the corresponding complete bipartite graph;
	
	\item[2) ] Determine the minimum, $g_{m}$, and maximum, $g_{M}$, values taken by $g$ ($g_{m} \le g \le g_{M}$) of the compact surface where $K_{m,n}$ may be embedded as a two-cell embedding;
	
	\item[3) ] Knowing $g_m$ and $g_M$, select a set of $(2g+1)$ symmetric points in the Poincaré disk for each value of the genus $g$ such that $g_m\leq g \leq g_M$, and establish the hyperelliptic curve;
	
	\item[4) ] From Whittaker's proposal, determine the uniformizing region of the hyperelliptic curve, or equivalently, the associated Fuchsian group, where the generators of the group are quotients of the linearly independent solutions from the FDE.
\end{itemize}

Step 4 is the most significant step since the generators of the Fuchsian group $\Gamma _0$ (elliptic transformations) establish the fundamental polygon associated with the given hyperelliptic function, from which the generators of a Fuchsian subgroup, $\Gamma_p \subset \Gamma_0$, are derived.

The algebraic manipulations and variable changes were exhaustively taken into consideration to fill in some essential steps left out of in obtaining the generators of the Fuchsian group associated with the uniformizing region of the given hyperelliptic curve. Because of the relevance and systematization of Whittaker's procedure in its completeness, we left to the Appendix for further considerations.

We have shown a variation of Whittaker's proposal for the construction of the signal sets. It consists of making use of the results coming from Whittaker's proposal, that is, where the vertices of the fundamental polygon, $\Gamma _0$, are inside the Poincaré disk as well as the fundamental polygon (associated with the generators of the subgroup $\Gamma_p$) which uniformizes the given hyperelliptic function. The variation was to take the previous vertices of the fundamental polygon, $\Gamma_0$, and place them at the boundary of the Poincaré disk by use of hyperbolic isometries. Hence, generating a new fundamental polygon (associated with the generators of $\Gamma_0^{\prime}$ as well as the fundamental polygon (associated with the generators of $\Gamma_p^{\prime}$) which uniformizes the given hyperelliptic function. The motivation of the latter case was to find the largest area (Voronoi region) associated with the decision region of a signal belonging to a signal set such that the performance of the digital transmission system in terms of the symbol error probability is the least possible.

Furthermore, we established a relation between the degree of the hyperelliptic function with the parameters of the tessellation $\{p,q\}$, leading to two cases as stated in Proposition \ref{prop_tesselacao2} and Proposition \ref{prop_tessel2}.

\newpage

\section{Appendix}

\subsection{Algebraic Procedure to Uniformize the Hyperelliptic Curve $y^2=z^{2g+1}+1$}

The aim of this Appendix is to consider a detailed description of Whittaker's proposal in order to fill in some essential steps left out of in obtaining the generators of the Fuchsian group associated with the uniformizing region of the given hyperelliptic curve. As it is well known, \cite{Whittaker1929}, the uniformizing variable $t$ for the hyperelliptic curve $y^2=(z- e_1)(z- e_2)...(z- e_{2g+2})= f(z)$ is such that the group of transformation consisting of the quotient of two linearly independent solutions of the second order differential equation

\begin{equation}\label{eq00}
\frac{d^2y}{dz^2}+\frac{3}{16}\left\{\sum_{r=1}^{2g+2} \frac{1}{(z-e_r)^2}+\frac{-(2g+2)z^{2g}+2gp_1z^{2g-1}+c_1z^{2g-2}+...+c_{2g-1}}{(z-e_1)(z-e_2)...(z-e_{2g+2})}\right\}y=0,
\end{equation}
where $p_1=\sum e_r$, and $c_1,...,c_{2g-1}$ are constants which depend on the roots of the hyperelliptic curve and are determined by the condition that the associated group is Fuchsian, \cite{Whittaker1898}, \cite{Mursi1930}, and \cite{Dhar1935} works which are closely followed. Thus, (\ref{eq00}) may be rewritten as
\begin{equation*}\label{001}
\frac{d^2y}{dz^2}+\frac{3}{16}\left\{ \left[\frac{f'(z)}{f(z)}\right]^2-\frac{2g+2}{2g+1}\frac{f''(z)}{f(z)}\right\}y=0.
\end{equation*}

The first approach in Whittaker's procedure is to arrive at the hypergeometric differential equation (HDE).  Thus, clever algebraic manipulations and proper variable changes are needed to obtain its solutions. And finally, to realize the algebraic development of the quotient of two linearly independent solutions.

Note that $f'(z)=(2g+1)z^{2g}$ and $f''(z)=(2g+1)(2g)z^{2g-1}$. Thus, (\ref{001}) is of the form
\begin{equation}\label{eqq11}
\frac{d^2y}{dz^2}+\frac{3}{16}\left[\frac{(2g+1)^2z^{4g}}{(z^{2g+1}+1)^2}-\frac{(2g+1)2g z^{2g-1}}{z^{2g+1}+1}\right]y=0.
\end{equation}

Consider $y(z)=u(z){f(z)}^{1/4}=u(z)(1+z^{2g+1})^{1/4}$. It follows that

$$y'(z)=u'(z)(1+z^{2g+1})^{1/4}+u(z)\frac{2g+1}{4}(1+z^{2g+1})^{-3/4}z^{2g},$$
and
\begin{eqnarray*}
	y''(z) &=& (1+z^{2g+1})^{1/4}u''(z)+ \frac{2g+1}{2}(1+z^{2g+1})^{-3/4}z^{2g}u'(z) \\
	& & + \left[-\frac{3(2g+1)^2}{16}(1+z^{2g+1})^{-7/4}z^{4g}+\frac{g(2g+1)}{2}(1+z^{2g+1})^{-3/4}z^{2g-1}\right]u(z).
\end{eqnarray*}

Substituting $y''(z)$, $y'(z)$ and $y(z)$ in (\ref{eqq11}) yields
\begin{eqnarray*}
	(1+z^{2g+1})^{1/4}u''(z) &+& \frac{2g+1}{2}(1+z^{2g+1})^{-3/4}z^{2g}u'(z) \\
	& +& \left[\frac{(2g+1)g}{2}-\frac{3(g+1)g}{4}\right](1+z^{2g+1})^{-3/4}z^{2g-1}u(z)=0.
\end{eqnarray*}

Multiplying the previous equation by $(1+z^{2g+1})^{3/4}$, we obtain
\begin{equation}\label{eqq22}
(1+z^{2g+1})u''(z)+ \frac{2g+1}{2}z^{2g}u'(z)+ \left[\frac{(2g+1)g}{2}-\frac{3(g+1)g}{4}\right]z^{2g-1}u(z)=0.
\end{equation}
Making a variable change $z$ to $s$ where $$s^2=1+z^{2g+1},$$ we see that $$z^{2g+1}=s^2-1 \Rightarrow z=(s^2-1)^{\frac{1}{2g+1}}.$$
And that, $$\frac{dz}{ds}=\frac{2}{2g+1}s(s^2-1)^{\frac{-2g}{2g+1}},$$ and
$$\frac{d^2z}{ds^2} = \frac{-8 g}{2g+1}s(s^2-1)^{\frac{-4g-1}{2g+1}}+ \frac{2}{2g+1}(s^2-1)^{\frac{-2g}{2g+1}}.$$

Thus,
\begin{itemize}
	\item $u(s)=u(z)$
	\item $u'(s)=u'(z)\frac{dz}{ds} \Rightarrow u'(z)=\frac{2g+1}{2}\frac{(s^2-1)^{\frac{2g}{2g+1}}}{s}u'(s)$
	\item $u''(s) =  u''(z)\left( \frac{dz}{ds}\right)^2+u'(z)\frac{d^2z}{ds^2} \rightarrow$
	\begin{eqnarray*}
		u''(z)& = & \frac{(2g+1)^2}{4}\frac{(s^2-1)^{\frac{4g}{2g+1}}}{s^2}u''(s)\\
		& & + \left[(2g+1)g\frac{(s^2-1)^{\frac{2g-1}{2g+1}}}{s}-\frac{(2g+1)^2}{4}\frac{(s^2-1)^{\frac{4g}{2g+1}}}{s^3}\right]u'(s)
	\end{eqnarray*}
\end{itemize}

Substituting $u''(z)$, $u'(z)$ and $u(z)$ in (\ref{eqq22}) leads to

\begin{eqnarray*}
	\frac{(2g+1)^2}{4}(s^2-1)^{\frac{4g}{2g+1}}u''(s)&+& \left[(2g+1)n s(s^2-1)^{\frac{2g-1}{2g+1}}\right]u'(s) \\
	& & +\left[\frac{(2g+1)g}{2}-\frac{3(g+1)g}{4}\right](s^2-1)^{\frac{2g-1}{2g+1}}u(s)=0.
\end{eqnarray*}

Multiplying the previous equation by $\frac{4}{2g+1}(s^2-1)^{\frac{-2g+1}{2g+1}}$, leads to

\begin{equation}\label{eqq33}
(2g+1)(s^2-1)u''(s)+ 4 g s u'(s)+\left[2g-\frac{3(g+1)g}{2g+1}\right]u(s)=0.
\end{equation}

Realizing a second variable change, namely, $s$ to $x$, where $$s=2x-1,$$ leads to $\frac{ds}{dx}=2$ and $\frac{d^2s}{dx^2}=0$.

Note that:
\begin{itemize}
	\item $u(x)=u(s)$
	\item $u'(x)=u'(s)\frac{ds}{dx} \Rightarrow u'(s)=\frac{1}{2}u'(x)$
	\item $u''(x)=u''(s)\left(\frac{ds}{dx}\right)^2+u'(s)\frac{d^2s}{dx^2} \Rightarrow u''(s)=\frac{1}{4}u''(x)$
	\item $s^2=(2x-1)^2=4x^2-4x+1 \Rightarrow s^2-1=4x(x-1)$
\end{itemize}

Substituting in (\ref{eqq33}) it follows that
\begin{equation}\label{eqhiperg}
x(1-x)u''(x)+\left[\frac{2g}{2g+1}-\frac{4g}{2g+1}x\right]u'(x)-\frac{g(g-1)}{(2g+1)^2}u(x)=0,
\end{equation}
is an hypergeometric differential equation (HDE).

The hypergeometric function associated to the HDE (\ref{eqhiperg}) is
$$F(\alpha, \beta; \gamma; x)=F\left(\frac{g-1}{2g+1}, \frac{g}{2g+1}; \frac{2g}{2g+1}; x\right).$$

To simplify the notation, we consider $a=\frac{1}{2g+1}$. Thus, $\alpha=(g-1)a$, $\beta=ga$ and $\gamma=2ga$.

Therefore, the hyperelliptic curve $y^2=z^{2g+1}+1$ whose associated FDE is given by (\ref{eqq11}) leads to the HDE (\ref{eqhiperg}) and the associated hypergeometric function is $$F(\alpha, \beta; \gamma; x)=F((g-1)a, ga; 2ga; x).$$

Next, the solutions of the HDE (\ref{eqhiperg}) are determined, for more information see \cite{Sotomayor1979}. Observe that the FDE has three singularities at $x=0$, $x=1$, and $x=\infty$.

The solutions in $x=0$ are
\begin{itemize}
	\item  $P=w_{10}(x)=F(\alpha, \beta; \gamma; x)=F((g-1)a, ga; 2ga; x),$
	
	\item  $Q=w_{20}(x)=x^a F(ga, (g+1)a; 2(g+1)a;x).$
\end{itemize}

The solutions in $x=1$ are
\begin{itemize}
	\item  $T=w_{11}(x)=F((g-1)a, ga; 2ga; 1-x),$
	
	\item  $U=w_{21}(x)=(1-x)^a F(ga, (g+1)a; 2(g+1)a;1-x).$
\end{itemize}

The solutions in $x=\infty$ are
\begin{itemize}
	\item  $R=w_{1\infty}(x)=x^{-(g-1)a}F((g-1)a, ga; 2ga; x^{-1}),$
	
	\item  $S=w_{2\infty}(x)=x^{-ga} F(ga, (g+1)a; 2(g+1)a;x^{-1}).$
\end{itemize}

Recall that a linear relation connects any of the previous three solutions with constant coefficients. Hence, without loss of generality, we may relate the solutions $P$, $R$ and $S$, that is,

\begin{eqnarray*}
	P &=&  \frac{\Gamma(2ga)\Gamma(a)}{\Gamma((g+1)a)\Gamma(ga)}(-x)^{-(g-1)a} F((g-1)a, ga; 2ga; x^{-1})\\
	& &   \\
	& & + \frac{\Gamma(2ga)\Gamma(-a)}{\Gamma((g-1)a)\Gamma(ga)}(-x)^{-ga} F((g+1)a, ga; 2(g+1)a;x^{-1}).
\end{eqnarray*}

Observe that:
\begin{itemize}
	\item $F(\alpha, \beta; \gamma; x)=F(\beta, \alpha; \gamma; x)$
	\item $(-x)^{-k}=x^{-k} e^{\pm k\pi i}$
\end{itemize}

Thus, $(-x)^{-(g-1)a}=x^{-(g-1)a} e^{\pm (g-1)a\pi i}$ and $(-x)^{-ga}=x^{-ga} e^{\pm ga\pi i}$.

Hence,
\begin{eqnarray*}
	P &=& F((g-1)a, ga; 2ga; x) \\
	&=& \frac{\Gamma(2ga)\Gamma(a)}{\Gamma((g+1)a)\Gamma(ga)}e^{\pm(g-1)a\pi i} x^{-(g-1)a} F(((g-1)a, ga; 2gna; x^{-1}) \\
	& &  +\frac{\Gamma(2ga)\Gamma(-a)}{\Gamma((g-1)a)\Gamma(ga)}e^{\pm ga\pi i}x^{-ga} F((g+1)a, ga; 2(g+1)a;x^{-1})\\
	&=&  \frac{\Gamma(2ga)\Gamma(a)}{\Gamma(gn+1)a)\Gamma(ga)}e^{\pm(g-1)a\pi i} R + \frac{\Gamma(2ga)\Gamma(-a)}{\Gamma((g-1)a)\Gamma(ga)}e^{\pm ga\pi i}S.
\end{eqnarray*}

Analogously, relating the solutions $Q$, $R$ and $S$ leads to
\begin{eqnarray*}
	Q &=& x^a F(ga, (g+1)a; 2(g+1)a;x) \\
	&  &  \\
	&=&  x^a \frac{\Gamma((2g+1)a)\Gamma(a)}{\Gamma((g+2)a)\Gamma((g+1)a)}(-x)^{-ga} F(ga, (g-1)a; 2ga; x^{-1})\\
	& &  \\
	& & + x^a \frac{\Gamma((2g+1)a)\Gamma(-a)}{\Gamma((g+1)a)\Gamma(ga)}(-x)^{-(g+1)a} F(ga, (g+1)a; 2(g+1)a;x^{-1}) \\
	& &  \\
	&=& \frac{\Gamma((2g+1)a)\Gamma(a)}{\Gamma((g+2)a)\Gamma((g+1)a)}e^{\pm ga\pi i}R +
	\frac{\Gamma((2g+1)a)\Gamma(-a)}{\Gamma((g+1)a)\Gamma(ga)}e^{\pm(g+1)a\pi i}S.
\end{eqnarray*}

Consider $t=\frac{Q}{P}$ in $i \infty$ and $t'=\frac{Q}{P}$ in $-i \infty$. Then,

$$t=\frac{\frac{\Gamma((2g+1)a)\Gamma(a)}{\Gamma((g+2)a)\Gamma((g+1)a)}e^{ga\pi i}R +
	\frac{\Gamma((2g+1)a)\Gamma(-a)}{\Gamma((g+1)a)\Gamma(ga)}e^{(g+1)a\pi i}S}{\frac{\Gamma(2ga)\Gamma(a)}{\Gamma((g+1)a)\Gamma(ga)}e^{(g-1)a\pi i} R +
	\frac{\Gamma(2ga)\Gamma(-a)}{\Gamma((g-1)a)\Gamma(ga)}e^{ga\pi i}S},$$

and

$$t'=\frac{\frac{\Gamma((2g+1)a)\Gamma(a)}{\Gamma((g+2)a)\Gamma((g+1)a)}e^{- ga\pi i}R +
	\frac{\Gamma((2g+1)a)\Gamma(-a)}{\Gamma((g+1)a)\Gamma(na)}e^{-(g+1)a\pi i}S}{\frac{\Gamma(2ga)\Gamma(a)}{\Gamma((g+1)a)\Gamma(ga)}e^{-(g-1)a\pi i} R +
	\frac{\Gamma(2ga)\Gamma(-a)}{\Gamma((g-1)a)\Gamma(na)}e^{- ga\pi i}S}.$$

Dividing the numerator and the denominator by $S$, yields

$$t=\frac{\frac{\Gamma((2g+1)a)\Gamma(a)}{\Gamma((g+2)a)\Gamma((g+1)a)}e^{ga\pi i}\frac{R}{S} +
	\frac{\Gamma((2g+1)a)\Gamma(-a)}{\Gamma((g+1)a)\Gamma(ga)}e^{(g+1)a\pi i}}{\frac{\Gamma(2ga)\Gamma(a)}{\Gamma((g+1)a)\Gamma(ga)}e^{(g-1)a\pi i} \frac{R}{S} +
	\frac{\Gamma(2ga)\Gamma(-a)}{\Gamma((g-1)a)\Gamma(ga)}e^{ga\pi i}},$$

and

$$t'=\frac{\frac{\Gamma((2g+1)a)\Gamma(a)}{\Gamma((g+2)a)\Gamma((g+1)a)}e^{- ga\pi i}\frac{R}{S} +
	\frac{\Gamma((2g+1)a)\Gamma(-a)}{\Gamma((g+1)a)\Gamma(ga)}e^{-(g+1)a\pi i}}{\frac{\Gamma(2ga)\Gamma(a)}{\Gamma((g+1)a)\Gamma(ga)}e^{-(g-1)a\pi i} \frac{R}{S}+
	\frac{\Gamma(2ga)\Gamma(-a)}{\Gamma((g-1)a)\Gamma(na)}e^{- ga\pi i}}.$$

The next step is to eliminate $\frac{R}{S}$. For that, we use the expression of $t$ to obtain $\frac{R}{S}$ and next, to obtain $\frac{S}{R}$. After multiplying and dividing $t'$ by $\frac{S}{R}$, leads to $t'$ as a function of $t$.

From the expression of $t$ it follows that

{\small
	\begin{eqnarray*}
		\frac{\Gamma(2ga)\Gamma(a)}{\Gamma((g+1)a)\Gamma(ga)}e^{(g-1)a\pi i} \frac{R}{S}t &+& \frac{\Gamma(2ga)\Gamma(-a)}{\Gamma((g-1)a)\Gamma(ga)}e^{ga\pi i} t  \\
		&= & \frac{\Gamma((2g+1)a)\Gamma(a)}{\Gamma((g+2)a)\Gamma((g+1)a)}e^{ga\pi i}\frac{R}{S} +\frac{\Gamma((2g+1)a)\Gamma(-a)}{\Gamma((g+1)a)\Gamma(ga)}e^{(g+1)a\pi i},
	\end{eqnarray*}
}
$$\frac{R}{S}= \frac{-\frac{\Gamma(2ga)\Gamma(-a)}{\Gamma((g-1)a)\Gamma(ga)}e^{ga\pi i}t + \frac{\Gamma((2g+1)a)\Gamma(-a)}{\Gamma((g+1)a)\Gamma(na)}e^{(g+1)a\pi i}}{\frac{\Gamma(2ga)\Gamma(a)}{\Gamma((g+1)a)\Gamma(ga)}e^{(g-1)a\pi i}t - \frac{\Gamma((2g+1)a)\Gamma(a)}{\Gamma((g+2)a)\Gamma((g+1)a)}e^{ga\pi i}}.
$$

Hence,
$$\frac{S}{R}= \frac{\frac{\Gamma(2ga)\Gamma(a)}{\Gamma((g+1)a)\Gamma(ga)}e^{(g-1)a\pi i}t - \frac{\Gamma((2g+1)a)\Gamma(a)}{\Gamma((g+2)a)\Gamma((g+1)a)}e^{ga\pi i}}{-\frac{\Gamma(2ga)\Gamma(-a)}{\Gamma((g-1)a)\Gamma(ga)}e^{ga\pi i}t + \frac{\Gamma((2g+1)a)\Gamma(-a)}{\Gamma((g+1)a)\Gamma(ga)}e^{(g+1)a\pi i}}.
$$

Finally, multiplying and dividing the expression of $t'$ by $\frac{S}{R}$ leads to
$$t'=\frac{\frac{\Gamma((2g+1)a)\Gamma(a)}{\Gamma((g+2)a)\Gamma((g+1)a)}e^{- ga\pi i} +
	\frac{\Gamma((2g+1)a)\Gamma(-a)}{\Gamma((g+1)a)\Gamma(ga)}e^{-(g+1)a\pi i}\frac{S}{R}}{\frac{\Gamma(2ga)\Gamma(a)}{\Gamma((g+1)a)\Gamma(ga)}e^{-(g-1)a\pi i} +
	\frac{\Gamma(2ga)\Gamma(-a)}{\Gamma((g-1)a)\Gamma(ga)}e^{- ga\pi i}\frac{S}{R}}.$$

Equivalently,
\begin{equation*}
t'=\frac{x_1 t+x_2}{x_3t+x_4},
\end{equation*}
where
$$x_1=\frac{\Gamma((2g+1)a)\Gamma(-a)}{\Gamma((g+1)a)\Gamma(ga)}\frac{\Gamma(2ga)\Gamma(a)}{\Gamma((g+1)a)\Gamma(ga)}
e^{-2a\pi i} - \frac{\Gamma((2g+1)a)\Gamma(a)}{\Gamma((g+2)a)\Gamma((g+1)a)}\frac{\Gamma(2ga)\Gamma(-a)}{\Gamma(ga)\Gamma((g-1)a)},$$

$$x_2=\frac{\Gamma((2g+1)a)\Gamma(a)}{\Gamma((g+2)a)\Gamma((g+1)a)}\frac{\Gamma((2g+1)a)\Gamma(-a)}{\Gamma((g+1)a)\Gamma(ga)}
\left[e^{a\pi i}-e^{-a\pi i}\right],$$

$$x_3=\frac{\Gamma(2ga)\Gamma(-a)}{\Gamma(ga)\Gamma((g-1)a)}\frac{\Gamma(2ga)\Gamma(a)}{\Gamma((g+1)a)\Gamma(ga)}
\left[e^{-a\pi i}-e^{a\pi i}\right],$$
and
$$x_4=\frac{\Gamma((2g+1)a)\Gamma(-a)}{\Gamma((g+1)a)\Gamma(ga)}\frac{\Gamma(2ga)\Gamma(a)}{\Gamma((g+1)a)\Gamma(ga)}
e^{2a\pi i} - \frac{\Gamma((2g+1)a)\Gamma(a)}{\Gamma((g+2)a)\Gamma((g+1)a)}\frac{\Gamma(2ga)\Gamma(-a)}{\Gamma(ga)\Gamma((g-1)a)}.$$
Note that
$$ e^{ai \pi}- e^{-a i\pi}= 2 i \sin(a\pi), \quad  e^{-a\pi i}-e^{a\pi i}=-2i\sin a\pi$$
and that
$$\Gamma(2ga)=(2ga-1)\Gamma((2ga-1)=-a\Gamma(-a),$$

$$\Gamma(2(g+1)a)=(2(g+1)a-1)\Gamma((2(g+1)a-1)=a\Gamma(a).$$

Hence,
\begin{eqnarray*}
	\Gamma(2ga)\Gamma(-a)\Gamma(2ga)\Gamma(a) &=&  \frac{a}{a}\Gamma(2ga)\Gamma(-a)\Gamma(2ga)\Gamma(a)\\
	&=& \frac{1}{a}[-a\Gamma(-a)]\Gamma(-a)\Gamma(2ga)a\Gamma(a)\\
	&=& -\Gamma(-a)\Gamma(-a)\Gamma(2ga)a\Gamma(2(g+1)a).
\end{eqnarray*}

Thus,
\begin{equation}\label{eq5}
t'= \frac{\left[\frac{\Gamma(2ga)\Gamma(a)}{\Gamma((g+1)a)\Gamma(ga)} e^{-2a\pi i} - \frac{\Gamma(2ga)\Gamma(a)}{\Gamma((g+2)a)\Gamma((g-1)a)}\right]t + \frac{\Gamma((2g+1)a)\Gamma(a)}{\Gamma((g+2)a)\Gamma((g+1)a)} 2i \sin a\pi}{\frac{\Gamma(2ga)\Gamma(-a)}{\Gamma(ga)\Gamma((g-1)a)} 2 i \sin a\pi t+ \left[\frac{\Gamma(2ga)\Gamma(a)}{\Gamma((g+1)a)\Gamma(ga)}e^{2a\pi i} - \frac{\Gamma(2ga)\Gamma(a)}{\Gamma((g+2)a)\Gamma((g-1)a)} \right] }.
\end{equation}

Recall that
\begin{itemize}
	\item $\Gamma(z)\Gamma(1-z)=\frac{\pi}{\sin \pi z}$
	\item $\Gamma(z)\Gamma(-z)=-\frac{\pi}{z\sin \pi z}$
	\item $\Gamma(z)=(z-1)\Gamma(z-1)$
\end{itemize}

Observe that
\begin{itemize}
	\item $\Gamma(2ga)=-a\Gamma(-a)$;
	\item $\Gamma((g+1)a)=-ga\Gamma(-ga)$;
	\item $\Gamma((g+2)a)=-(g-1)a\Gamma(-(g-1)a)$.
\end{itemize}

From this, we have
\begin{itemize}
	\item $\Gamma(2ga)\Gamma(a)=\frac{\pi}{\sin a\pi}$
	\item $\Gamma((g+1)a)\Gamma(na)=\frac{\pi}{\sin ga\pi}$
	\item $\Gamma((g+2)a)\Gamma((g-1)a)=\frac{\pi}{\sin (g-1)a\pi}$
\end{itemize}

When substituting these functions in (\ref{eq5}), yields.

\begin{equation}\label{eq6}
t'=\frac{\left[ \frac{\sin ga\pi}{\sin a\pi} e^{-2a\pi i} -\frac{\sin (g-1)a\pi}{\sin a\pi}\right]t + \frac{\Gamma((2g+1)a)\Gamma(a)}{\Gamma(g+2)a)\Gamma((g+1)a)} 2i \sin a\pi}
{\frac{\Gamma(2ga)\Gamma(-a)}{\Gamma(ga)\Gamma((g-1)a)} 2 i \sin a\pi t+
	\left[ \frac{\sin ga\pi}{\sin a\pi} e^{2a\pi i} -\frac{\sin (g-1)a\pi}{\sin a\pi}\right]}.
\end{equation}

Let $t_1=\frac{\Gamma(2ga)\Gamma(-a)}{\Gamma(ga)\Gamma((g-1)a)} t$, then

\[
t=\frac{\Gamma(ga)\Gamma((g-1)a)}{\Gamma(2ga)\Gamma(-a)}t_1 \quad \mbox{and}\quad  t'=\frac{\Gamma(na)\Gamma((g-1)a)}{\Gamma(2ga)\Gamma(-a)}t'_1.
\]

Substituting this last $t'$ in (\ref{eq6}), we have

$$\frac{\Gamma(ga)\Gamma((g-1)a)}{\Gamma(2ga)\Gamma(-a)}t'_1=\frac{\left[ \frac{\sin ga\pi}{\sin a\pi} e^{-2a\pi i} -\frac{\sin (g-1)a\pi}{\sin a\pi}\right]\frac{\Gamma(ga)\Gamma((g-1)a)}{\Gamma(2ga)\Gamma(-a)}t_1 + \frac{\Gamma((2g+1)a)\Gamma(a)}{\Gamma((g+2)a)\Gamma((g+1)a)} 2i \sin a\pi}
{2 i \sin a\pi t_1+
	\left[ \frac{\sin ga\pi}{\sin a\pi} e^{2a\pi i} -\frac{\sin (g-1)a\pi}{\sin a\pi}\right]},$$

\begin{eqnarray*}
	&&\frac{\Gamma(ga)\Gamma((g-1)a)}{\Gamma(2ga)\Gamma(-a)}t'_1=\\
	&&\frac{\Gamma(ga)\Gamma((g-1)a)}{\Gamma(2ga)\Gamma(-a)}\frac{\left[ \frac{\sin na\pi}{\sin a\pi} e^{-2a\pi i} -\frac{\sin (g-1)a\pi}{\sin a\pi}\right]t_1 + \frac{\Gamma((2g+1)a)\Gamma(a)}{\Gamma((g+2)a)\Gamma((g+1)a)} \frac{\Gamma(ga)\Gamma((g-1)a)}{\Gamma(2ga)\Gamma(-a)}2i \sin a\pi}
	{2 i \sin a\pi t_1+
		\left[ \frac{\sin ga\pi}{\sin a\pi} e^{2a\pi i} -\frac{\sin (g-1)a\pi}{\sin a\pi}\right]}.
\end{eqnarray*}

Note that
\[
\frac{\Gamma((2g+1)a)\Gamma(a)}{\Gamma((g+2)a)\Gamma((g+1)a)}\frac{\Gamma(2ga)\Gamma(-a)}{\Gamma(ga)\Gamma((g-1)a)}=
\frac{\Gamma((2g+1)a)\Gamma(-a)}{\Gamma((g+2)a)\Gamma((g-1)a)}\frac{\Gamma(2ga)\Gamma(a)}{\Gamma(ga)\Gamma((g+1)a)}
\]
\[
= -\frac{\sin (g-1)a\pi}{\sin a\pi}\frac{\sin ga\pi}{\sin a\pi},
\]
and
\[
\sin(g-1)a\pi \sin ga\pi=\frac{1}{2}\left[\cos -a\pi -\cos (2g-1)a\pi\right]=\frac{1}{2}\left[\cos a\pi +\cos 2a\pi\right].
\]

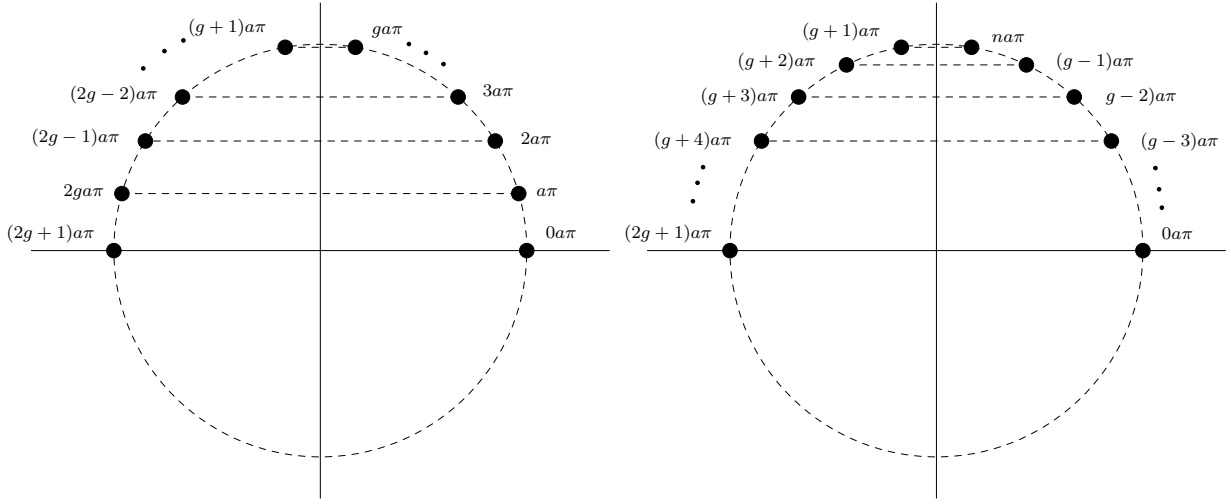
\begin{figure}[!h]
	\resizebox{1.0\linewidth}{!}{
		\begin{tikzpicture}[line cap=round,line join=round,>=triangle 45,x=1.0cm,y=1.0cm]
		\begin{scope}[xshift=3cm, yshift=1cm, scale=3.25]
		\draw [dashed] (0.,0.) circle (1.cm);
		\draw [dashed] (-0.9610482568468216,0.2763806216283727)-- (0.9610482568468216,0.27638062162837257);
		\draw [dashed] (0.8472674358151735,0.5311665390507772)-- (-0.8472674358151735,0.5311665390507773);
		\draw [dashed] (-0.6675167294205647,0.7445947998365773)-- (0.6675167294205648,0.7445947998365772);
		\draw [dashed] (-0.17013542250147953,0.9854206908778823)-- (0.17013542250147973,0.9854206908778823);
		\draw  (0.,1.2)-- (0.,-1.2);
		\draw  (-1.4,0.)-- (1.4,0.);
		\begin{scriptsize}
		\draw [fill=black] (0.9610482568468216,0.27638062162837257) circle (1.pt);
		\draw[color=black] (1.01391856543036305,0.283173764186558952) node[right] {$a\pi$};
		\draw [fill=black] (-0.9610482568468216,0.2763806216283727) circle (1.pt);
		\draw[color=black] (-1.013489608579578,0.285952668852151093) node[left] {$2ga \pi$};
		\draw [fill=black] (0.8472674358151735,0.5311665390507772) circle (1.pt);
		\draw[color=black] (0.9387070703707962056,0.5449604778692946) node[right] {$2a \pi$};
		\draw [fill=black] (0.6675167294205648,0.7445947998365772) circle (1.pt);
		\draw[color=black] (0.753741107359623069,0.71587200696961629) node[above right] {$3a \pi$};
		\draw [fill=black] (0.17013542250147973,0.9854206908778823) circle (1.pt);
		\draw[color=black] (0.32555555735828099,1.002153677786699246) node[above] {$ga \pi$};
		\draw [fill=black] (-0.8472674358151735,0.5311665390507773) circle (1.pt);
		\draw[color=black] (-0.9380411850533198808,0.54676197505604467) node[left] {$(2g-1)a \pi$};
		\draw [fill=black] (-0.6675167294205647,0.7445947998365773) circle (1.pt);
		\draw[color=black] (-0.753762988368082,0.75872700696961629) node[left] {$(2g-2)a \pi$};
		\draw [fill=black] (-0.17013542250147953,0.9854206908778823) circle (1.pt);
		\draw[color=black] (-0.22673581055169852,1.002903677786699246) node[above left] {$(g+1)a \pi$};
		\draw [fill=black] (-1.,0.) circle (1.0pt);
		\draw[color=black] (-1.05929076773284797,0.08107194262930811154) node[left] {$(2g+1)a \pi$};
		\draw [fill=black] (1.,0.) circle (1.0pt);
		\draw[color=black] (1.0591275559839795046,0.087194262930811154) node[right] {$0 a \pi$};
		\draw [fill=black] (0.5117406277093084,0.95845100824285071) circle (0.25pt);
		\draw [fill=black] (0.5965267999494078,0.90398857570925464) circle (0.25pt);
		\draw [fill=black] (0.4290008216000074,1.00107989992229897) circle (0.25pt);
		\draw [fill=black] (-0.66311270956563831,1.018741341557976) circle (0.25pt);
		\draw [fill=black] (-0.85752657836520837,0.88309608920253543) circle (0.25pt);
		\draw [fill=black] (-0.7550171812940842,0.96711227848277189) circle (0.25pt);
		\end{scriptsize}
		\end{scope}
		\end{tikzpicture}
		\begin{tikzpicture}[line cap=round,line join=round,>=triangle 45,x=1.0cm,y=1.0cm]
		\begin{scope}[xshift=3cm, yshift=1cm, scale=3.25]
		\draw [dashed] (0.,0.) circle (1.cm);
		\draw [dashed] (0.8472674358151735,0.5311665390507772)-- (-0.8472674358151735,0.5311665390507773);
		\draw [dashed] (-0.6675167294205647,0.7445947998365773)-- (0.6675167294205648,0.7445947998365772);
		\draw [dashed] (-0.17013542250147953,0.9854206908778823)-- (0.17013542250147973,0.9854206908778823);
		\draw (0.,1.2)-- (0.,-1.2);
		\draw  (-1.4,0.)-- (1.4,0.);
		\draw [dashed] (-0.43579187619076554,0.9000474657739628)-- (0.4357918761907654,0.9000474657739628);
		\begin{scriptsize}
		\draw [fill=black] (0.8472674358151735,0.5311665390507772) circle (1pt);
		\draw[color=black] (0.95606195322110052,0.530449604778692946) node[right] {$(g-3)a \pi$};
		\draw [fill=black] (0.6675167294205648,0.7445947998365772) circle (1pt);
		\draw[color=black] (0.78776598973771065,0.747700696961629) node[right] {$g-2)a \pi$};
		\draw [fill=black] (0.17013542250147973,0.9854206908778823) circle (1pt);
		\draw[color=black] (0.23555555735828099,0.98) node[above right] {$na \pi$};
		\draw [fill=black] (-0.8472674358151735,0.5311665390507773) circle (1pt);
		\draw[color=black] (-0.9545723436735805,0.53291735073070871) node[left] {$(g+4)a \pi$};
		\draw [fill=black] (-0.6675167294205647,0.7445947998365773) circle (1pt);
		\draw[color=black] (-0.728671502787217819,0.747700696961629) node[left] {$(g+3)a \pi$};
		\draw [fill=black] (-0.17013542250147953,0.9854206908778823) circle (1pt);
		\draw[color=black] (-0.23673581055169852,0.9853677786699246) node[above left] {$(g+1)a \pi$};
		\draw [fill=black] (-1.,0.) circle (1pt);
		\draw[color=black] (-1.051929076773284797,0.0817194262930811154) node[left] {$(2g+1)a \pi$};
		\draw [fill=black] (1.,0.) circle (1pt);
		\draw[color=black] (1.0596318973154205,0.0817194262930811154) node[right] {$0a \pi$};
		\draw [fill=black] (1.091856543036305,0.20804682158279156) circle (0.25pt);
		\draw [fill=black] (1.0606195322110052,0.3999314205274224) circle (0.25pt);
		\draw [fill=black] (1.0784692526826052,0.29729547225471287) circle (0.25pt);
		\draw [fill=black] (-1.1795203869747797,0.23928384931796404) circle (0.25pt);
		\draw [fill=black] (-1.1572082363852798,0.3285324999898854) circle (0.25pt);
		\draw [fill=black] (-1.13043365567788,0.4043938530610185) circle (0.25pt);
		\draw [fill=black] (-0.43579187619076554,0.9000474657739628) circle (1pt);
		\draw[color=black] (-0.54895663427088834,0.91621169733593445) node[left] {$(g+2)a \pi$};
		\draw [fill=black] (0.4357918761907654,0.9000474657739628) circle (1pt);
		\draw[color=black] (0.546059000821600008,0.8390287296757585563) node[above right] {$(g-1)a \pi$};
		\end{scriptsize}
		\end{scope}
		\end{tikzpicture}
	}
	\caption{Equivalent angles for hyperelliptic curves of degree $2g+1$}
	\label{fig_cap2_angulo_equiv_2nmais1}
\end{figure}

Fig. \ref{fig_cap2_angulo_equiv_2nmais1} illustrates the equivalent angles. Such equivalence are necessary to perform some simplifications.

Thus, $$t'_1=\frac{\left[ \frac{\sin ga\pi}{\sin a\pi} e^{-2a\pi i} -\frac{\sin (g-1)a\pi}{\sin a\pi}\right]t_1 -i \frac{\cos a\pi +\cos 2a\pi}{\sin a\pi}}
{(2 i \sin a\pi) t_1+ \left[ \frac{\sin ga\pi}{\sin a\pi} e^{2a\pi i} -\frac{\sin (g-1)a\pi}{\sin a\pi}\right]}.$$
Observe that

\begin{itemize}
	\item $\frac{\sin ga\pi}{\sin a\pi} e^{-2a\pi i} -\frac{\sin (g-1)a\pi}{\sin a\pi}= 2\cos a\pi e^{-(g+1)a\pi i}$\\
	\listpart{In fact,
	\begin{eqnarray*}\label{contas_equiv_pag15}
		& &\frac{\sin ga\pi}{\sin a\pi} e^{-2a\pi i} -\frac{\sin (g-1)a\pi}{\sin a\pi}\\
		&=& \frac{\sin ga\pi e^{-2a\pi i} -\sin (g-1)a\pi}{\sin a\pi} \\
		&=& \frac{\sin ga\pi \left(\cos 2a\pi- i \sin 2a\pi\right) -\sin (g-1)a\pi}{\sin a\pi} \\
		&=& \frac{\sin ga\pi \cos 2a\pi- i \sin ga\pi \sin 2a\pi  -\sin (g-1)a\pi}{\sin a\pi} \\
		&=& \frac{\sin (-g)a\pi \cos (2g-1)a\pi- i \sin ga\pi \sin (2g-1)a\pi  -\sin (g-1)a\pi}{\sin a\pi} \\
		&=& \frac{\frac{1}{2}\left[\sin ((2g-1)a\pi-ga\pi)-\sin ((2g-1)a\pi+ga\pi)\right]}{\sin a\pi} \\
		& & - \frac{i\frac{1}{2}\left[\cos ((2g-1)a\pi- ga\pi)-\cos ((2g-1)a\pi+ ga\pi)\right]-\sin (g-1)a\pi}{\sin a\pi} \\
		&=& \frac{\frac{1}{2} \left[\sin (g-1)a\pi-\sin (3g-1)a\pi\right] -i\frac{1}{2}\left[\cos (g-1)a\pi-\cos (3g-1)a\pi\right]-\sin (g-1)a\pi}{\sin a\pi} \\
		&=& \frac{-\frac{1}{2} \left[\sin (3g-1)a\pi+\sin (g-1)a\pi\right] +i\frac{1}{2}\left[\cos (3g-1)a\pi-\cos (g-1)a\pi\right]}{\sin a\pi} \\
		&=& \frac{-\sin (2g-1)a\pi \cos (-g)a\pi +i \sin (2g-1)a\pi \sin (-g)a\pi}{\sin a\pi} \\
		&=& \frac{-\sin (2g-1)a\pi \left[\cos ga\pi +i \sin ga\pi\right]}{\sin a\pi} \\
		&=& \frac{-\sin (2g-1)a\pi \left[-\cos (g+1)a\pi +i \sin (g+1)a\pi\right]}{\sin a\pi} \\
		&=& \frac{\sin (2g-1)a\pi \left[\cos (g+1)a\pi -i \sin (g+1)a\pi\right]}{\sin a\pi} \\
		&=& \frac{\sin (2g-1)a\pi e^{-(g+1)a\pi i}}{\sin a\pi} \\
		&=& \frac{\sin 2a\pi e^{-(g+1)a\pi i}}{\sin a\pi} \\
		&=& \frac{2\sin a\pi \cos a\pi e^{-(g+1)a\pi i}}{\sin a\pi} \\
		&=& 2 \cos a\pi e^{-(g+1)a\pi i}.
	\end{eqnarray*}}
	\item $\frac{\sin ga\pi}{\sin a\pi} e^{2a\pi i} -\frac{\sin (g-1)a\pi}{\sin a\pi}= 2\cos a\pi e^{(g+1)a\pi i}$\\
	\listpart{In fact,
	\begin{eqnarray*}\label{contas_equiv_pag16}
		& &\frac{\sin ga\pi}{\sin a\pi} e^{2a\pi i} -\frac{\sin (g-1)a\pi}{\sin a\pi}\\
		&=& \frac{\sin ga\pi e^{2a\pi i} -\sin (g-1)a\pi}{\sin a\pi} \\
		&=& \frac{\sin ga\pi \left(\cos 2a\pi+ i \sin 2a\pi\right) -\sin (g-1)a\pi}{\sin a\pi} \\
		&=& \frac{\sin ga\pi \cos 2a\pi+ i \sin ga\pi \sin 2a\pi  -\sin (g-1)a\pi}{\sin a\pi} \\
		&=& \frac{-\sin ga\pi \cos (2g-1)a\pi+ i \sin ga\pi \sin (2g-1)a\pi  -\sin (g-1)a\pi}{\sin a\pi} \\
		&=& \frac{-\frac{1}{2}\left[\sin ((-g)a\pi+a\pi)+\sin (3ga \pi-a\pi)\right]}{\sin a\pi} \\
		& & + \frac{i\frac{1}{2}\left[\cos ((-g)a\pi+ a\pi)-\cos (3ga\pi- a\pi)\right]-\sin (g-1)a\pi}{\sin a\pi} \\
		&=& \frac{-\frac{1}{2} \left[\sin (-(g-1))a\pi+\sin (3g-1)a\pi\right]}{\sin a\pi}\\
		& & +\frac{i\frac{1}{2}\left[\cos (-(g-1))a\pi-\cos (3g-1)a\pi\right]-\sin (g-1)a\pi}{\sin a\pi} \\
		&=& \frac{-\frac{1}{2} \left[\sin (3g-1)a\pi+\sin (g-1)a\pi\right] -i\frac{1}{2}\left[\cos (3g-1)a\pi-\cos (g-1)a\pi\right]}{\sin a\pi} \\
		&=& \frac{-\sin (2g-1)a\pi \cos ga\pi +i \sin (2g-1)a\pi \sin ga\pi}{\sin a\pi} \\
		&=& \frac{-\sin (2g-1)a\pi \left[\cos ga\pi -i \sin ga\pi\right]}{\sin a\pi} \\
		&=& \frac{-\sin 2a\pi \left[-\cos (g+1)a\pi -i \sin (g+1)a\pi\right]}{\sin a\pi} \\
		&=& \frac{\sin 2a\pi \left[\cos (g+1)a\pi +i \sin (g+1)a\pi\right]}{\sin a\pi} \\
		&=& \frac{2 \sin a\pi \cos a\pi e^{(g+1)a\pi i}}{\sin a\pi} \\
		&=& 2 \cos a\pi e^{(g+1)a\pi i}.
	\end{eqnarray*}}
\end{itemize}

From this, it follows that
\begin{equation}\label{eqa}
t'_1=\frac{2\cos a\pi e^{-(g+1)a\pi i}t_1 -i \frac{\cos a\pi +\cos 2a\pi}{\sin a\pi}}
{2 i \sin a\pi t_1+ 2\cos a\pi e^{(g+1)a\pi i}}.
\end{equation}

This is the transformation when $x$ passes from $i\infty$ to $-i\infty$ rounding the singularity $x=1$.

Now, for $x$ rounding $x=0$, we have to use the solutions $T$ and $U$ which holds at $x=1$. Similarly, to the previous case, as $x$ passes from $i\infty$ to $-i\infty$ rounding $x=0$, the transformation is given by

\begin{equation}\label{eqaw}
t'_1 = e^{2a\pi i} t_1.
\end{equation}

Thus, by substituting the transformations $$t_1=it_2 \cot a\pi e^{ga\pi i}$$ and $$t_2=t_3\sec a\pi \sqrt{(\cos a\pi + \cos 2a\pi )/2}$$ in (\ref{eqaw}) and in (\ref{eqa}), we have
\begin{eqnarray*}
	i \cot a\pi e^{ga\pi i} t'_2 &=& \frac{2\cos a\pi e^{-(g+1)a\pi i}i \cot a\pi e^{ga\pi i} t_2 -i \frac{\cos a\pi +\cos 2a\pi}{\sin a\pi}}{2 i \sin a\pi i \cot a\pi e^{ga\pi i} t_2+ 2\cos a\pi e^{(g+1)a\pi i}} \\
	&=& \frac{2i\cos^2 a\pi/\sin a\pi \left[ e^{-a\pi i}i t_2 - \frac{\cos a\pi +\cos 2a\pi}{2\cos^2 a\pi}\right]}
	{-2\cos a\pi e^{(g+1)a\pi}\left[e^{-a\pi i} t_2-1\right]} \\
	&=&- i \cot a\pi e^{-(g+1)a\pi i} \frac{e^{-a\pi i}i t_2 - \frac{\cos a\pi +\cos 2a\pi}{2\cos^2 a\pi}}
	{e^{-a\pi i} t_2-1} \\
	&=& i \cot a\pi e^{ga\pi i} \frac{e^{-a\pi i}i t_2 - \frac{\cos a\pi +\cos 2a\pi}{2\cos^2 a\pi}}
	{e^{-a\pi i} t_2-1}
\end{eqnarray*}

Hence,
\begin{equation}\label{eqbb}
t'_2=\frac{e^{-a\pi i}i t_2 - \frac{\cos a\pi +\cos 2a\pi}{2\cos^2 a\pi}}
{e^{-a\pi i} t_2-1}
\end{equation}

By changing the variables $$t_2=\sqrt{\frac{(\cos a\pi +\cos 2a\pi)}{2\cos^2 a\pi}}t_3 \quad \mbox{and} \quad t'_2=\sqrt{\frac{(\cos a\pi +\cos 2a\pi)}{2\cos^2 a\pi}}t'_3,$$in (\ref{eqbb}), we obtain
\begin{eqnarray*}
	\sqrt{\frac{\cos a\pi +\cos 2a\pi}{2\cos^2 a\pi}}t'_3 &=& \frac{e^{-a\pi i}i \sqrt{\frac{\cos a\pi +\cos 2a\pi}{2\cos^2 a\pi}}t_3 - \frac{\cos a\pi +\cos 2a\pi}{2\cos^2 a\pi}}{e^{-a\pi i}\sqrt{\frac{\cos a\pi +\cos 2a\pi}{2\cos^2 a\pi}}t_3 -1}\\
	&=&\sqrt{\frac{\cos a\pi +\cos 2a\pi}{2\cos^2 a\pi}}  \frac{e^{-a\pi i} (\frac{(\cos a\pi +\cos 2a\pi}{2\cos^2 a\pi})^{-1/2})t_3 - 1}{e^{-a\pi i} t_3-(\frac{\cos a\pi +\cos 2a\pi}{2\cos^2 a\pi})^{-1/2}}
\end{eqnarray*}
Therefore, in place of (\ref{eqa}) and (\ref{eqaw}), we have
\[
t'_3=\frac{e^{-a\pi i} (\frac{\cos a\pi +\cos 2a\pi}{2\cos^2 a\pi})^{-1/2}t_3 - 1}{e^{-a\pi i} t_3 -(\frac{\cos a\pi +\cos 2a\pi}{2\cos^2 a\pi})^{-1/2}},
\]
and
\[
t'_3 = e^{(2\pi ai)}t_3.
\]

Thus, these transformations lead to a transformation which is the quotient of the two solutions of (\ref{eqhiperg}) when $z$ passes successively by the two circuits, namely, from the infinite to the singularity $z_1=e^{a\pi i}$, round it, and returning to infinite, and from the infinite to the singularity $z_2=e^{3a\pi i}$, round it, and returning to infinite. By using some algebraic manipulations, we end up with the Mobius transformations given by

\[
S_j(t)=\frac{\left(2\cos (a\pi) -1\right)^{-1/2} t-e^{\frac{1}{2}(4k+1)a\pi i}}{e^{-\frac{1}{2}(4k+1)a\pi i}t-\left(2\cos (a\pi) -1\right)^{-1/2}},
\]
or in the matrix form by
\[
S_j=\left(
\begin{array}{ll}
\left(2\cos (a\pi) -1\right)^{-1/2} & -e^{\frac{1}{2}(4k+1)a\pi i} \\
&    \\
e^{-\frac{1}{2}(4k+1)a\pi i} & -\left(2\cos (a\pi) -1\right)^{-1/2} \\
\end{array}
\right),
\]
where $a=1/(2g+1)$, $j=1,\ldots, (2g+1)$ and $k=0,1, 2, \cdots , 2g$ are the generators of the Fuchsian group whose fundamental polygon $\mathcal{P}_{(2g+1)}$ has $(2g+1)$ sides, for $g\ge 2$.

The region which uniformizes the curve $y^2 = z^{(2g+1)} +1$ consists of the juxtaposition of two of these polygons giving rise to a polygon $\mathcal{P}_{4g}$ with $4g$ sides. The $2g$ generators of the Fuchsian group $\Gamma_{4g}$ are given by $S_2S_1$, $S_3S_1$, $S_4S_1$, $\cdots $ $S_{2g+1}S_1$.

\end{document}